\documentclass[useAMS,usenatbib]{mn2e}
\usepackage[utf8]{inputenc}
\usepackage{graphicx,epsfig,ragged2e}
\usepackage{amsmath,amssymb}
\usepackage{xcolor}
\usepackage{url}
\title[Giant radio halo in ACT-CL J0256.5+0006]{A giant radio halo in a low-mass SZ-selected galaxy cluster: ACT-CL J0256.5+0006\vspace{-0.5cm}}

\author[K. Knowles et al.]{\Large K.~Knowles$^{1,}$\thanks{E-mail: kendaknowles.astro@gmail.com}, 
H.T.~Intema$^2$, 
A.J.~Baker$^3$, 
V.~Bharadwaj$^{1}$,
J.R.~Bond$^4$, 
C.~Cress$^{5,6}$, 
N.~Gupta$^7$, 
\newauthor \Large A.~Hajian$^{4}$, 
M.~Hilton$^{1}$, 
A.D.~Hincks$^{8}$ 
R.~Hlozek$^{9}$, 
J.P.~Hughes$^{3,}$\thanks{Visiting Astronomer, Gemini South Observatory}, 
R.R.~Lindner$^{3,10}$, 
T.A.~Marriage$^{11}$, 
\newauthor \Large F.~Menanteau$^{12,13}$, 
K.~Moodley$^1$, 
M.D.~Niemack$^{14}$, 
E.D.~Reese$^{15}$, 
J.~Sievers$^{16,17}$, 
C.~Sif\'{o}n$^{18}$, 
\newauthor \Large R.~Srianand$^7$, 
E.J.~Wollack$^{19}$
\\
\footnotesize $^1$ Astrophysics \& Cosmology Research Unit, School of Mathematics, Statistics and Computer Science, University of KwaZulu-Natal,\\ Durban 4041, South Africa\\
$^2$ National Radio Astronomy Observatory, 1003 Lopezville Road, Socorro, NM 87801, USA\\
$^3$ Department of Physics and Astronomy, Rutgers, The State University of New Jersey, 136 Frelinghuysen Road, Piscataway,\\ NJ 08854-8019, USA\\
$^4$ Canadian Institute for Theoretical Astrophysics, University of Toronto, Toronto, ON M5S 3H8, Canada\\
$^5$ Centre for High Performance Computing, CSIR Campus, 15 Lower Hope Rd, Rosebank, Cape Town, South Africa\\
$^{6}$ Physics Department, University of the Western Cape, Modderdam Rd, Bellville, 7535\\
$^7$ IUCAA, Post Bag 4, Ganeshkhind, Pune 411007, India \\
$^{8}$ Department of Physics and Astronomy, University of British Columbia, 6224 Agricultural Rd., Vancouver BC V6T 1Z1, Canada\\
$^9$ Department of Astrophysical Sciences, Peyton Hall, Princeton University, Princeton, NJ 08544, USA\\
$^{10}$ Department of Astronomy, The University of Wisconsin-Madison, 475 N. Charter Street, Madison, WI 53706-1582, USA\\
$^{11}$ Department of Physics and Astronomy, The Johns Hopkins University, 3400 N. Charles St., Baltimore, MD 21218-2686, USA\\
$^{12}$ National Center for Supercomputing Applications, University of Illinois at Urbana-Champaign, 1205 W. Clark St., Urbana, IL 61801, USA\\
$^{13}$ Department of Astronomy, University of Illinois at Urbana-Champaign, W. Green Street, Urbana, IL 61801, USA\\
$^{14}$ Department of Physics, Cornell University, Ithaca, NY 14853, USA\\
$^{15}$ Department of Physics, Astronomy, and Engineering, Moorpark College, 7075 Campus Rd., Moorpark, CA  93021, USA\\
$^{16}$ Astrophysics \& Cosmology Research Unit, School of Chemistry \& Physics, University of KwaZulu-Natal, Durban 4041, South Africa\\
$^{16}$ National Institute for Theoretical Physics (NITheP), University of KwaZulu-Natal, Private Bag X54001, Durban 4000, South Africa\\
$^{18}$ Leiden Observatory, Leiden University, PO Box 9513, NL2300 RA Leiden, Netherlands\\
$^{19}$ NASA/Goddard Space Flight Center, Observational Cosmology Laboratory, 8800 Greenbelt Rd, Greenbelt, MD 20771, USA}

\newcommand{\degrees}{$^\circ$}
\newcommand{\per}{$^{-1}$}
\date{Accepted XXX. Received YYY; in original form ZZZ}

\pubyear{2016}

\begin{document}
\label{firstpage}
\pagerange{\pageref{firstpage}--\pageref{lastpage}}

\maketitle

\begin{abstract}
We present the detection of a giant radio halo (GRH) in the Sunyaev-Zel'dovich (SZ)-selected merging galaxy cluster ACT-CL J0256.5+0006 ($z = 0.363$), observed with the Giant Metrewave Radio Telescope at 325 MHz and 610 MHz. We find this cluster to host a faint ($S_{610} = 5.6 \pm 1.4$ mJy) radio halo with an angular extent of 2.6 arcmin, corresponding to 0.8 Mpc at the cluster redshift, qualifying it as a GRH. J0256 is one of the lowest-mass systems, $M_{\rm 500,SZ} = (5.0 \pm 1.2) \times 10^{14} M_\odot$, found to host a GRH.  We measure the GRH at lower significance at 325 MHz ($S_{325} = 10.3 \pm 5.3$ mJy), obtaining a spectral index measurement of $\alpha^{610}_{325} = 1.0^{+0.7}_{-0.9}$. This result is consistent with the mean spectral index of the population of typical radio halos, $\alpha = 1.2 \pm 0.2$. Adopting the latter value, we determine a 1.4 GHz radio power of $P_{1.4\text{GHz}} = (1.0 \pm 0.3) \times 10^{24}$ W Hz\per, placing this cluster within the scatter of known scaling relations. 
Various lines of evidence, including the ICM morphology, suggest that ACT-CL J0256.5+0006 is composed of two subclusters. We determine a merger mass ratio of 7:4, and a line-of-sight velocity difference of $v_\perp = 1880 \pm 210$ km s{\per}. We construct a simple merger model to infer relevant time-scales in the merger. From its location on the $P_{\rm 1.4GHz}$--$L_{\rm X}$ scaling relation, we infer that we observe ACT-CL J0256.5+0006 just before first core crossing.
\end{abstract}

\clearpage
\section{Introduction}
Multiwavelength observations of galaxy clusters provide a wealth of information about the physics of the intracluster medium (ICM) and its relationship with cluster galaxies. The optical and X-ray bands have historically been used to identify merger activity via optical substructure \citep{CarterMetcalfe.1980.OptGCMorph, GellerBeers.1982.OptSubstruct, RheeKatgert.1987.OptSubstructAbellGCs, DresslerShectman.1988.DSTest, Rhee.1991.OptSubstructAbellGCs, WenHan.2013.optsubstruct} and morphological parameters determined from X-ray images \citep{Mohr.1993.XSubstruct, Jeltema.2005.MorphParams, OHara.2006.XMorphParams, Santos.2008.MorphParams}. In the last decade, a link has been found between a cluster's merger status and the presence of large-scale diffuse synchrotron emission \citep[see][and references therein]{BrunettiJones.2014.Review}. This cluster-scale radio emission, dubbed a giant radio halo (GRH) if $\sim$Mpc in size, exhibits a steep spectrum and has no obvious link to the individual cluster galaxies \
citep{Buote.2001.GRH, FerettiGiovannini.2008.GRH, Ferrari.2008.GRHReview, Feretti.2012.Review}. Radio halos (RHs) appear to trace the non-thermal ICM and typically have spectral indices of $\alpha \sim$ 1.1--1.5. However, ultra-steep spectrum radio halos (USSRHs, $\alpha \sim$ 1.6--1.9), presumably associated with more pronounced synchrotron ageing, have also been detected within the population \citep{Brunetti.2008.USSRH, Dallacasa.2009.USSRH, Venturi.2013.EGRHS1}.

The existence of USSRHs is predicted by one of the current leading theories for the origin of RHs \citep{Brunetti.2008.USSRH}, namely the \textit{turbulent re-acceleration} model in which the synchrotron emission is powered by turbulence generated during cluster mergers \citep{Brunetti.2001.RH_Coma, Petrosian.2001.coma, BrunettiLazarian.2011.primary, Beresnyak.2013.MHDTurb}. In this model one expects an USSRH to be seen when the turbulent energy in the cluster has decreased sufficiently for it to be less efficient in accelerating high energy electrons in the cluster. This scenario can also explain the observed bimodality in scaling relations between the 1.4 GHz RH power and thermal cluster properties, in which clusters are observed to be either radio loud or radio quiet. This dichotomy has been observed in cluster samples selected via X-ray luminosity \citep{Brunetti.2007.CRandGRH, Cassano.2008.RevStats} and the Sunyaev-Zel'dovich (SZ) effect \citep{SunyaevZeldovich.1972.SZeffect}, although it is less 
pronounced in the latter case \citep{ SommerBasu.2014.SZvsX}. In practice, one anticipates a population of clusters in transition between these two states that will have intermediate radio power.

The observed bimodality was initially thought to be due in part to selection effects in the cluster sample \citep{Basu.2012.SZbimodality}, but this has since been ruled out \citep{SommerBasu.2014.SZvsX, Cuciti.2015.SZwRHs}. A more likely reason is a physical effect related to the cluster evolutionary state. Magnetohydrodynamic (MHD) simulations by \citet{Donnert.2013.MHDSims} show that a RH is a transient phenomenon that exhibits a rise and fall in radio halo emission over the course of a merger. This evolutionary model suggests that for a merging cluster, the observable diffuse radio emission depends strongly on the phase of the merger in which the cluster is being observed, which likely contributes to the scatter in the observed $P_{\rm 1.4GHz}$ scaling relations with thermal cluster properties.

Moreover, one would expect to find two separate types of systems that populate the intermediate region of radio power: late-stage mergers with old RHs that are in the process of switching off, and early-stage mergers in which the radio halo emission has recently switched on but not yet reached its maximum radio power. The former scenario is a possible explanation for some of the observed USSRHs, which are starting to fill in the region between the correlation and upper limits. Clusters that are in the early stages of merging would also be interesting systems to identify and study as they would complete the evolutionary picture; however due to their expected low radio power, they are potentially more difficult to detect.

In line with the above, \citet{Cassano.2010.GRHMergerConn} find that the observed dichotomy is strongly related to cluster dynamical state, with morphologically disturbed systems hosting RHs. However, several RH non-detections in merging clusters are seemingly incongruent with this trend (A141, A2631, MACSJ2228: \citealt{Cassano.2010.GRHMergerConn}; A119: \citealt{GiovanniniFeretti.2000.GRHDets}; and A2146: \citealt{Russell.2011.A2146}). In the case of A2146, \citet{Russell.2011.A2146} postulate that the lack of a RH in this strongly-merging system is due to the relatively low mass of the cluster. They estimate a radio power upper limit more than an order of magnitude below the correlation. Low-mass systems are expected to generate less turbulent energy during their mergers, yielding weaker synchrotron emission, and hence RHs that are too faint to observe with current telescopes. The era of LOFAR \citep{Vermeulen.2012.LOFAR}, SKA precursors such as MeerKAT \citep{BoothJonas.2012.MeerKAT} and ASKAP \citep{
DeBoer.2009.ASKAP}, and the SKA itself \citep{Taylor.2013.SKA} will bring with it highly sensitive observations of these systems, and should reveal the underlying RH emission.

In this paper we present the detection of a GRH in a low-mass system that we argue is in the early stages of merging. As discussed, such early-stage merging systems are interesting because they allow us to probe the full evolutionary cycle of GRHs and are expected to fill in the intermediate region in radio halo power.

The paper is organised as follows: we present existing multiwavelength data on ACT-CL J0256.5+0006 in Section \ref{sec:J0256}, and we describe the radio observations and data reduction process in Section \ref{sec:radioobs}, with the radio results presented in Section \ref{sec:radioresults}. X-ray and optical morphological analyses are discussed in Sections \ref{sec:xray} and \ref{sec:optical}, respectively. We construct a model for the merger geometry in Section \ref{sec:mergergeom} and infer merger time-scales from this model in Section \ref{sec:timescales}. We conclude with a discussion in Section \ref{sec:conclusion}. In this paper we adopt a $\Lambda$CDM flat cosmology with $H_0 = 70$ km s\per Mpc\per, $\Omega_m$ = 0.27 and $\Omega_\Lambda$ = 0.73. In this cosmology, at the redshift of our cluster ($z$=0.363), one arcminute corresponds to 305.8 kpc. We assume $S_\nu \propto \nu^{-\alpha}$ throughout the paper, where $S_\nu$ is the flux density at frequency $\nu$ and $\alpha$ is the spectral index. Colour 
versions of all figures are available in the online journal.

\section{ACT-CL J0256.5+0006}
\label{sec:J0256}
ACT-CL J0256.5+0006 (hereafter J0256) lies at $z$=0.363 and was detected by the Atacama Cosmology Telescope \citep[ACT;][]{Kosowsky.2006.ACT} equatorial SZ cluster survey with a 148 GHz decrement signal-to-noise ratio of 5.4 \citep{Hasselfield.2013.ACTE}. It was first identified in \textit{ROSAT} PSPC data and is included in the Bright \textsc{sharc} catalog \citep[RX J0256.5+0006;][]{Burke.1997.SHARC}. \citet{Majerowicz.2004.J0256} identify J0256 as undergoing a major merger based on observations carried out with \emph{XMM-Newton}.

In the following sub-sections we describe the existing multiwavelength data for J0256 in the X-ray (\textit{XMM-Newton}), optical (Gemini), millimetre (ACT), and radio (VLA) bands. The relevant cluster properties are given in Table \ref{table:j0256}. 

\begin{table}
 \caption{Published properties of J0256.}
 \label{table:j0256}
 \centering
  \begin{tabular}[h]{ll}
    \hline
    R.A. (hh mm ss.s) & 02 56 33.0 $^a$\\
    Dec. (dd mm ss.s) & +00 06 26.3 $^a$\\ 
    redshift & 0.363 $^b$\\
    $L_{\rm 500,X} $ ($10^{44} \text{ ergs s}^{-1}$) & 3.01 {$\pm$} 0.36 $^c$ \\
    $Y_{500}$ ($10^{-4} \text{ arcmin}^2$) & 3.4 {$\pm$} 1.0 $^d$ \\
    $M_{\rm 500,X} $ ($10^{14} \text{ M}_\odot$) & $5.2^{+1.1}_{-0.9}$ $^e$\\
    $M_{\rm 500,SZ} $ ($10^{14} \text{ M}_\odot$) & 5.0 {$\pm$} 1.2 $^d$\\
    \hline
  \end{tabular}
  
  \justify
  \begin{itemize}
   \item [$^a$] R.A. and Dec. (J2000) of the SZ peak of the cluster, with an \indent\indent astrometric accuracy of 5-10\arcsec.
   \item [$^b$] \citet{Menanteau.2013.ACTE}
   \item [$^c$] Integrated 0.1\textendash2.4 keV X-ray luminosity using the \indent\indent spectrum of \citet{Majerowicz.2004.J0256}, corrected for \indent\indent the cosmology adopted in this paper.
   \item [$^d$] Integrated Compton $y$-parameter and B12 SZ mass from \indent\indent \citet{Hasselfield.2013.ACTE}.
   \item [$^e$] Total mass for the main cluster component using $\beta$-model fit \indent\indent``a'' for the NE region \citep{Majerowicz.2004.J0256}.
  \end{itemize}
\end{table}

\subsection{X-ray}
\label{sec:arch-Xray}
\citet{Majerowicz.2004.J0256}, hereafter M04, carry out a comprehensive X-ray study of J0256 based on their 25.3 ks \emph{XMM-Newton} observations (obs ID: 005602301)\footnote{The \textit{XMM-Newton} observations include European Photon Imaging Camera (EPIC) data from the two MOS (Metal Oxide Semi-conductor) CCD arrays and the pn CCD array.}. The X-ray image shows two components in the direction of the cluster: a bright main component and a less luminous structure to the west. To investigate whether these are physically connected or serendipitously aligned, M04 fit an elliptical $\beta$-model to the hot gas distribution of the main component, excluding point sources and the western component. After subtraction of the best-fit model from the data, the residuals reveal that the western component is a small galaxy cluster exhibiting a comet-like morphology, with the tail to the west (see Figure 2 in M04). This orientation indicates that gas in the subcluster is undergoing ram pressure stripping as it interacts 
with the main cluster component. Based on the orientation of the subcluster isophots away from the main component and numerical simulations by \citet{RickerSarazin.2001.xraymergers}, M04 conclude that the subcluster has not yet passed through the main cluster centre and thus that J0256 is in the pre-core crossing stage of its merger.

For the full cluster, M04 determine a temperature of $T = 4.9^{+0.5}_{-0.4}$ keV within $\sim0.8R_{500}$ and a bolometric X-ray luminosity\footnote{Corrected for the cosmology used in this paper.} of $L_{\rm X} = (7.88 \pm 0.53) \times 10^{44}$ erg s\per, which is over-luminous compared to the $L_{\rm X}$--$T$ relation measured by \citet{ArnaudEvrard.1999.LXTRel}. M04 conclude that this discrepancy between observed and predicted luminosity, coupled with their evidence for ram pressure stripping of the subcluster, suggests J0256 is not in dynamical equilibrium. Using \textsc{xspec}\footnote{\url{https://heasarc.gsfc.nasa.gov/xanadu/xspec/}} to model the M04 spectrum using the cosmology in this paper, we determined a 0.1-2.4 keV band-limited luminosity of $L_{\rm 500,X} = (3.01 \pm 0.36) \times 10^{44}$ erg s{\per}, incorporating a conservative 10\% uncertainty due to the spectrum being extracted within $\sim$0.8$R_{500}$. 

From $\beta$-model fitting, M04 calculate an X-ray mass for the main cluster component of $M_{500,X} = 3.7^{+0.8}_{-0.6} \times 10^{14} M_\odot$, which is equivalent to $M_{500,X} = 5.2^{+1.1}_{-0.9} \times 10^{14} M_\odot$ using the cosmology in this paper. The M04 $M_{200}$ mass for the main cluster component is $M_{200} \sim 5.9 \times 10^{14} M_\odot$.  Using count rates in the residual map in the region of the subcluster and translating this into a luminosity, they estimate the $M_{200}$ mass of the subcluster to be between 1-2 $\times 10^{14} M_\odot$ and determine a merger mass ratio of $\sim$ 3:1. However, this calculation requires several broad assumptions due to a lack of ancillary data, making the result somewhat uncertain.

\subsection{Millimetre}
\label{sec:arch-Micro}
Wide area, untargeted SZ surveys detect large numbers of galaxy clusters via inverse Compton scattering of cosmic microwave background (CMB) photons by electrons within the hot ICM, which causes a distortion of the CMB spectrum in the direction of clusters. ACT is a 6 m telescope that observes the millimetre sky with arcminute resolution \citep{Swetz.2011.ACT}. Between 2008 and 2011, ACT surveyed a 455 deg$^2$ strip centred at $\delta$ = -55{\degrees}, as well as a 504 deg$^2$ strip centred at $\delta$ = 0{\degrees} overlapping the Sloan Digital Sky Survey (SDSS) Stripe 82 \citep{Marriage.2011.ACTS, Hasselfield.2013.ACTE}, at 148, 218, and 277 GHz. ACT has detected over ninety clusters via the SZ effect.

J0256 was identified in the ACT equatorial 148 GHz map, with a decrement signal-to-noise ratio of 5.4 for a filter scale of $\theta_{500}$ = 7.06{\arcmin} \citep[see][hereafter H13, for details]{Hasselfield.2013.ACTE}. H13 investigated prescriptions for the pressure profile used to obtain a {$Y_{500}$\textemdash$M_{500}$} scaling relation, where $Y_{500}$ is the integrated Compton parameter. H13 investigated several profiles computed from simulations \citep[e.g.,][]{Battaglia.2012.B12profile} or empirical models \citep[e.g.,][]{Arnaud.2010.UPP}, leading to a SZ mass range of 2.9 $\times 10^{14} M_\odot < M_{500} < 7.5 \times 10^{14} M_\odot$ for J0256, taking into account the range of uncertainties on all mass estimates. The pressure profile from \citet{Battaglia.2012.B12profile} is currently preferred, and in this paper we use the corresponding SZ mass estimate of $M_{\rm 500,SZ} = (5.0 \pm 1.2) \times 10^{14} M_\odot$. 

\begin{figure*}
 \centering
 \includegraphics[width=0.9\textwidth, clip]{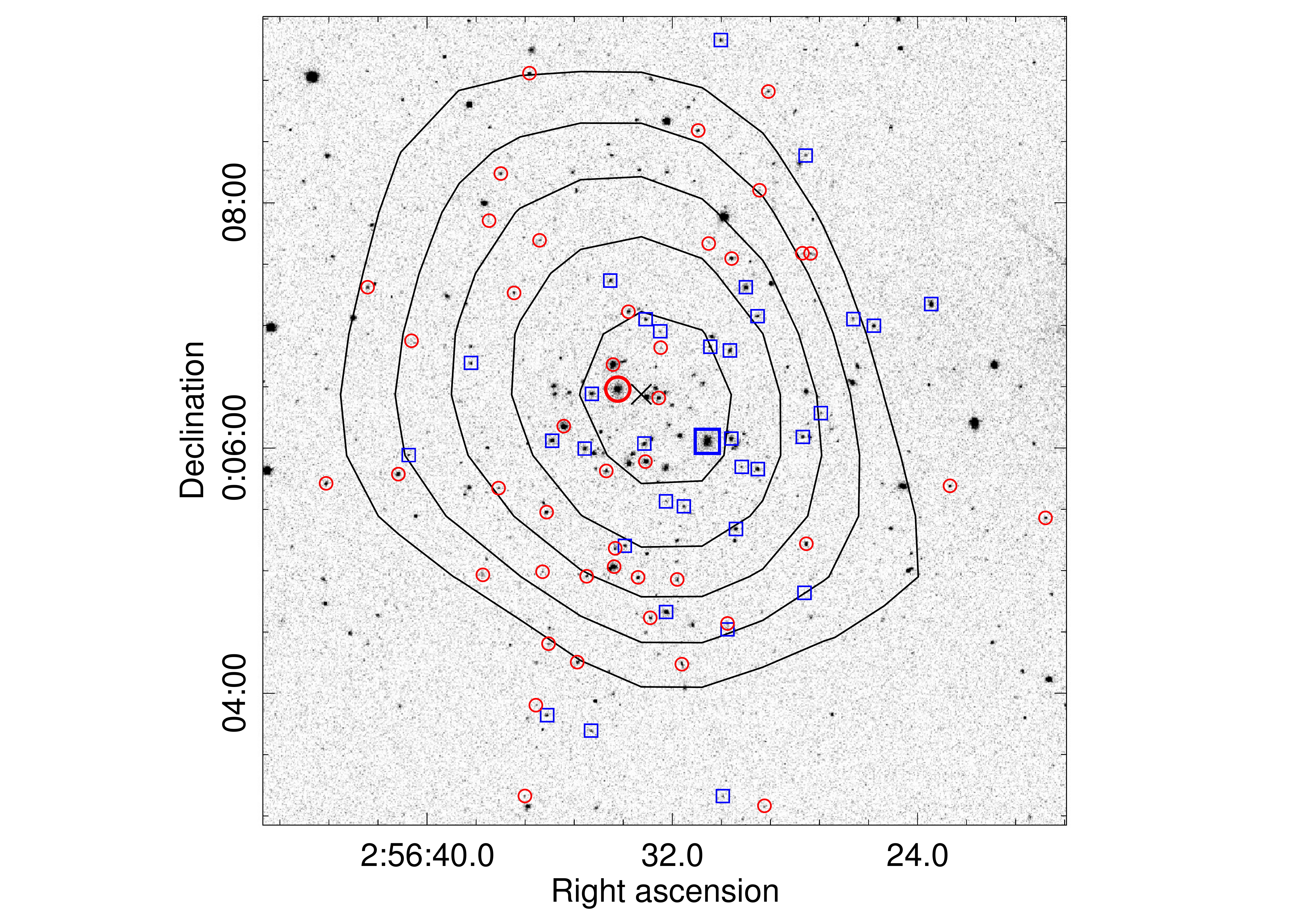}
 \caption{Cluster member galaxies with spectroscopic redshifts from Gemini identified on an SDSS $r$-band image. Blue boxes (red circles) denote members with higher (lower) redshifts than the systemic cluster redshift of $z$ = 0.363. Large, bold symbols mark the BCGs of both kinematic components. The 148 GHz Compton $y$ SZ contours are superposed. The contours start at a level of $2.0 \times 10^{-5}$, increasing towards the centre in steps of $1.25 \times 10^{-5}$. The black X marks the cluster SZ peak.}
 \label{fig:opticalmembers}
\end{figure*}

\subsection{Optical}
\label{sec:arch-Opt}
The ACT collaboration has completed spectroscopic observations of J0256 using Gemini and identified 78 cluster members \citep{Sifon.2015}. This distribution of spectroscopically confirmed cluster members is $\sim$85\% complete within $R_{200}$, up to an $r$-band magnitude limit of 22. Using this redshift information, we can estimate an independent dynamical mass and re-examine the merger geometry proposed by M04 (see Section \ref{sec:optical} below). The cluster members are shown in Figure \ref{fig:opticalmembers} where red circles (blue boxes) denote members that are at lower (higher) redshifts than the cluster redshift of $z$ = 0.363. We identify these two sets of galaxies as separate kinematic components (see Section \ref{sec:optical} below), each of which has a brightest cluster galaxy (BCG) that is indicated by a large, bold symbol. If the cluster is not in the core passage phase of its merger, the superposition of the two populations in the plane of the sky indicates that the merger is occurring at 
least partially along the line-of-sight.

\subsection{Radio}
\label{sec:arch-Radio}
\begin{figure*}
 \centering
 \includegraphics[width=0.32\textwidth]{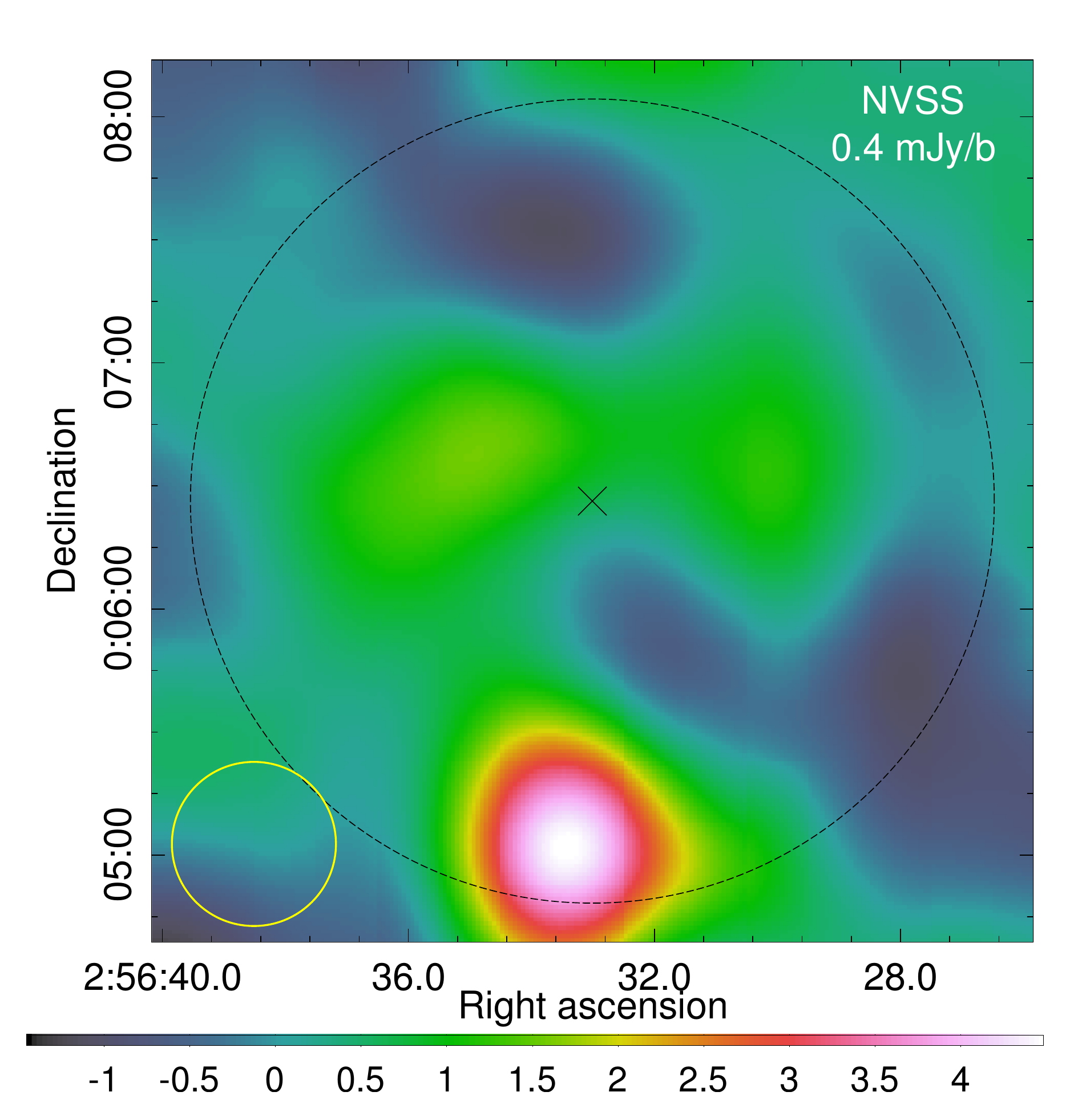}
 \includegraphics[width=0.32\textwidth]{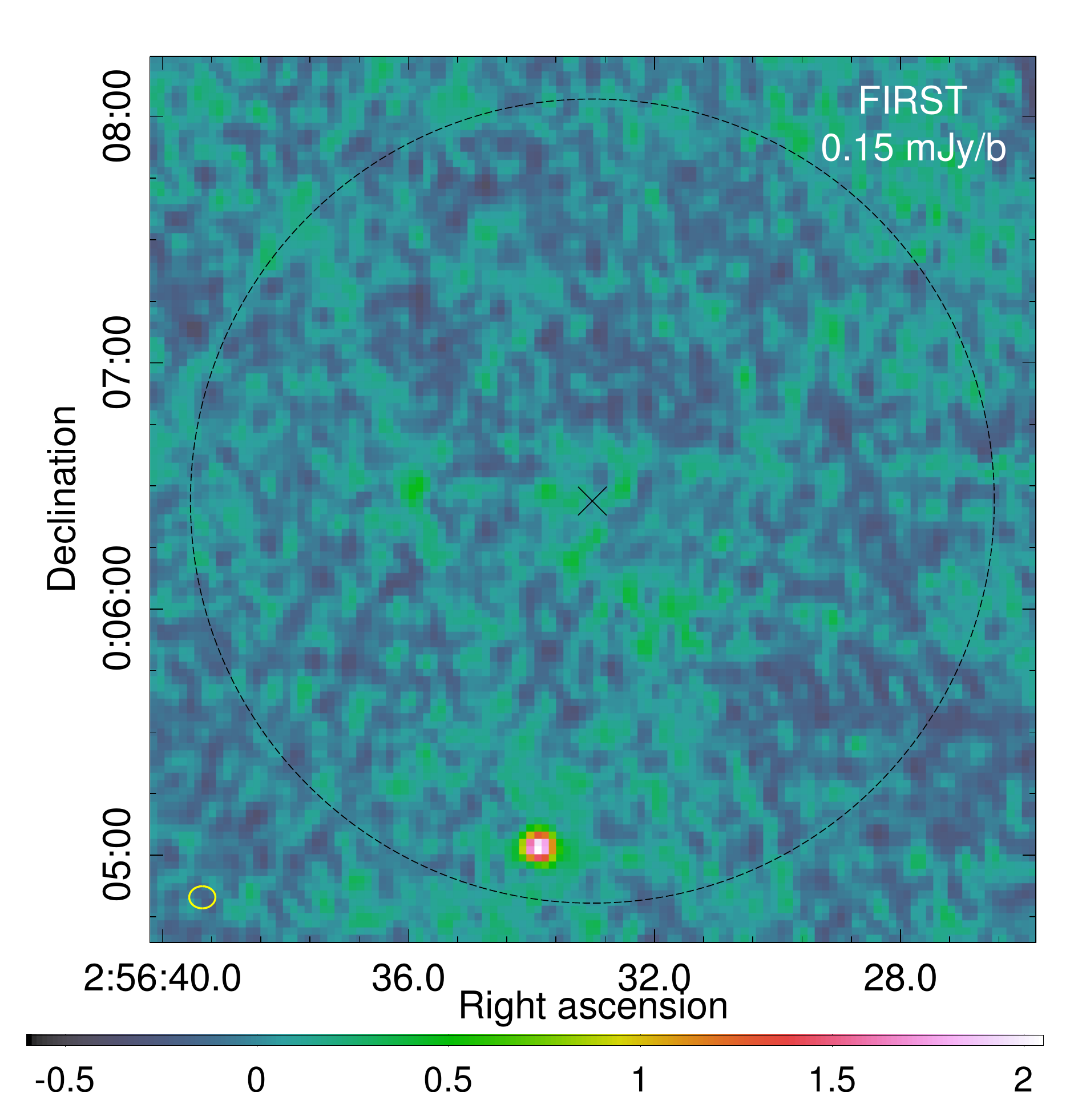}
 \includegraphics[width=0.32\textwidth]{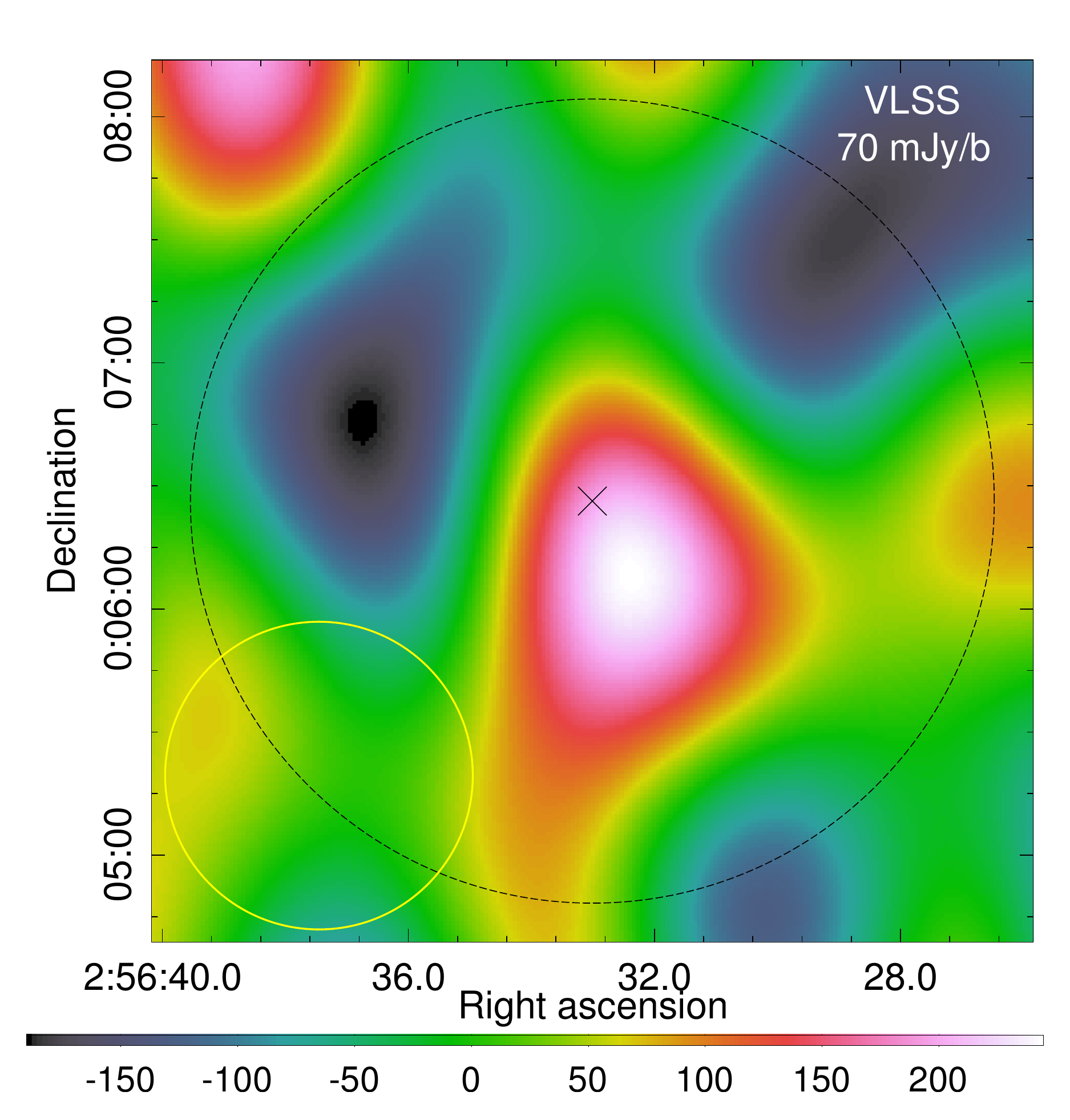}
 \caption{Postage stamp images of the J0256 cluster region at 1.4 GHz from NVSS (left) and FIRST (middle), and at 74 MHz from VLSS (right). The dashed black circle denotes $R_{500}$ centred on the SZ peak, which is marked by a black X. The image resolutions, from left to right, are 40{\arcsec} $\times$ 40\arcsec, 6.4{\arcsec} $\times$ 5.4\arcsec, and 75{\arcsec} $\times 75$\arcsec. The rms is given in the upper right corner and the beam is indicated by the yellow ellipse at lower left in each image. The colour scales are all in units of mJy beam\per.}
 \label{fig:archRadio}
\end{figure*}

J0256 has been mapped at 1.4 GHz in the NRAO VLA Sky Survey \citep[NVSS;][]{Condon.1998.NVSS} and the Faint Images of the Radio Sky at Twenty-Centimetres \citep[FIRST;][]{Becker.1995.FIRST} survey and at 74 MHz in the VLA Low-Frequency Sky Survey \citep[VLSS;][]{Cohen.2007.VLSS}. Figure \ref{fig:archRadio} shows the cluster region in each of the three sky surveys. Only one point source is detected in the 1.4 GHz survey data at R.A. and Dec. (J2000) of 02h56m34s and +00d065m03. Its NVSS and FIRST fluxes are 4.8 {$\pm$} 0.4 mJy and 3.66 {$\pm$} 0.27 mJy, respectively. This source is not detected in the VLSS data; however, there is a source 1.16{\arcmin} away, closer to the SZ peak of the cluster, detected 3$\sigma$ above the map noise. The rms and resolution of each image is given in the caption for Figure \ref{fig:archRadio}.

\begin{table*}
 \centering
 \caption{GMRT observations. }
  \begin{tabular}[t]{ccccccccc}
    \hline\hline
    Frequency$^a$ & Observing date & On-source & Integration & Bandwidth$^b$ & $\theta_{\rm synth}$, p.a.$^c$ & rms noise$^c$ & HPBW & ${\theta_{\rm max}}^d$\\
    (MHz) &  & time (hrs) & time (s) & (MHz) & (\arcsec$\times$\arcsec, \degrees) & ($\umu$Jy beam\per) & (arcmin) & (arcmin)\\
    \hline
    610 & Aug 2012 & 7.5 & 16.1 & 29.1 & 5.7 $\times$ 4.3, 71.3 & 26 & 43 $\pm$ 3 & 17 \\
    325 & Jul 2014 & 6.5 & 8.1 & 31.2 & 9.8 $\times$ 8.2, 76.1 & 72 & 81 $\pm$ 4 & 32 \\
    \hline
  \end{tabular}
  
  \justify
  $^a$ Observing frequency. \\$^b$ Bandwidth remaining after flagging. \\$^c$ Synthesised beam and rms noise of the full-resolution images, where p.a. denotes the beam position angle.\\$^d$ Maximum recovered scale.
  \label{table:obsstats}
\end{table*}

\section{New Radio Observations}
\label{sec:radioobs}

We observed J0256 with the Giant Metrewave Radio Telescope (GMRT) as part of an ongoing project involving the radio follow-up of ACT equatorial clusters. Initial observations were carried out for 10 hours at 610 MHz in August 2012 (PI: Knowles), using a 33 MHz bandwidth split into 256 channels and a 16s integration time. The data were acquired in the polarization channels RR and LL, and the total on-source time was 7.5 hrs. Flux and bandpass calibrator 3C48 was observed at the beginning, middle, and end of the observing block. This source was also used to estimate the instrument's antenna gains and ionospheric phase calibration which in turn were used to correct observations of the target field. A second set of 8-hour observations was carried out at 325 MHz on the GMRT using Director's Discretionary Time (PI: Knowles) in July 2014. This dataset has a central frequency of 323 MHz with a total bandwidth of 33 MHz made up of 256 channels and an integration time of 8s. The total on-source time was 6.5 hrs. As 
with the 610 MHz observations, 3C48 was used as the sole calibrator. Observational details are given in Table \ref{table:obsstats}. The pointing centre for both sets of observations was the same and was defined to be that of the SZ peak, given in Table \ref{table:j0256}.

The 610 MHz and 325 MHz data were subjected to the same calibration procedure, which is based on AIPS (NRAO Astronomical Image Processing System), SPAM \citep{Intema.2009.SPAM}, and Obit \citep{Cotton.2008.Obit} tools. The main calibration steps are outlined here. First, strong radio frequency interference (RFI) is removed by statistical outlier flagging tools. As a compromise between imaging speed and spectral resolution losses due to bandwidth smearing, the datasets are then averaged down to 24 channels. Phase calibration starts from a model derived from the VLSS \citep{Cohen.2007.VLSS} and the NVSS \citep{Condon.1998.NVSS}, followed by a succession of self-calibration loops. To compensate for the non-coplanarity of the array, we use the polyhedron (facet-based) wide-field imaging technique available in AIPS. We perform several rounds of imaging and self-calibration, inspecting the residual visibilities for more accurate removal of low-level RFI using Obit. To correct for ionospheric effects, we then apply 
SPAM calibration and imaging. The presence of strong sources in the field of view enables one to derive direction-dependent (DD) gains for each source and to use these gains to fit a time variable phase screen over the entire array. The phase screen was used during imaging to correct the full field of view for ionospheric phase effects.

As J0256 lies at close to zero declination, bright sources in the field are subject to strong north-south sidelobes that interfere with emission in the cluster region. To reduce the impact of these bright sources during further imaging, we modeled and subtracted all sources in the field outside of a 13 arcminute radius centred on the cluster, leaving a dataset with only the inner portion of the field. This edited $uv$-dataset was then imported into the Common Astronomy Software Applications package  \citep[CASA;][]{McMullin.2007.CASA} for imaging.

For each dataset we created several target field images, all with Briggs robust $R$ = 0 weighting \citep{Briggs.1995.BRW}. We first made full resolution (FR) images, shown in Figures \ref{fig:610BEST} (610 MHz) and \ref{fig:330BEST} (325 MHz) in the Appendix, using all of the $uv$-data, cleaning until the residuals were noise-like. We then created high-resolution (HR) images in the following way. As the 610 MHz data have more long baselines than the 325 MHz data, we matched the $uv$-coverage of the two datasets by selecting a $uv$-range from 4 k$\lambda$ ($\sim$52\arcsec) to 30 k$\lambda$ ($\sim$6\arcsec), and imaging using a 25 k$\lambda$ outer taper. The HR images were cleaned until their residuals showed no indication of emission in the cluster region. The clean components from the HR images were used as compact source models and were subtracted from the $uv$-data to create a point source-subtracted datasets. Using these datasets, we imaged at full resolution (PSSUB-FR) to visually check that the point 
source subtraction was successful. 610 MHz HR and PSSUB-FR images of the cluster region are compared in the left and right panels of Figure \ref{fig:ptsrcsubB+A} respectively. The PSSUB-FR image shows no visual indication of residual emission from the compact sources; however, we nevertheless investigate contamination from the source removal process in Section \ref{subsec:ptsrccontam}. Once satisfied, we re-imaged with a $uv$-cut of $<$ 4 k$\lambda$ and an outer taper of 3 k$\lambda$ to gain sensitivity to diffuse emission on scales of ~1 Mpc, creating point source subtracted, low-resolution (PSSUB-LR) images. We convolved each PSSUB-LR image with a 1{\arcmin} Gaussian, providing better sensitivity to extended features while retaining useful data, to create our final smoothed, point source subtracted, low-resolution (LR) maps shown in Figures \ref{fig:610LR} (610 MHz) and \ref{fig:330LR} (325 MHz) in the Appendix. The final LR 610 MHz (325 MHz) map has a maximum angular resolution of 17{\arcmin} (32{\arcmin})
. A summary of the different images created is given in Table \ref{table:radioimages}.

\begin{table}
 \centering
 \caption{Properties of the different radio images created. Values in brackets are for the 325 MHz images when different from the corresponding 610 MHz images. }
 \begin{tabular}{cccccccc}
  \hline\hline
  Image ID & ${\theta_{\rm min}}^\star$ & ${\theta_{\rm max}}^\dagger$ & Point sources \\
   & (arcmin) & (arcmin) & removed\\
  \hline
  FR & 0.08 (0.13) & 17 (32) & No\\
  HR & 0.13 & 0.86 & No \\
  PSSUB-FR & 0.08 (0.13) & 17 (32) & Yes\\
  PSSUB-LR & 0.84 & 17 (32) & Yes\\
  LR$^\ddagger$ & 1.30 (1.26) & 17 (32) & Yes\\
  \hline
 \end{tabular}
 
 \justify
 $^\star$ The highest resolution available, defined by the synthesised beam.\\
 $^\dagger$ The largest scale to which the image is sensitive, defined by the shortest baseline/$uv$-wavelength.\\
 $^\ddagger$ PSSUB-LR convolved with a 1{\arcmin} Gaussian. 1{\arcmin} corresponds to $\sim$3.5 k$\lambda$.
 \label{table:radioimages}
\end{table}

\begin{table*}
 \centering
 \caption{Properties of cluster region radio sources. Source labels are shown in the left panel of Figure \ref{fig:GMRT610}. The given R.A. and Dec. are for the peak source emission in the 610 MHz map. Flux errors include 10\% measurement uncertainties. The uncertainties on $\alpha$ are determined via numerical methods, as described in Section \ref{subsec:radioptsrcs}.}
 \begin{tabular}{cccccccl}
  \hline\hline
  Source & RA & DEC & Type$^a$ & $S_{610}$ & $S_{325}$ & $\alpha^b$ & Notes\\
   & (hms) & (dms) & & (mJy) & (mJy) \\
  \hline
  S1 & 02 56 35.5 & 00 06 11.0 & C & 0.56 {$\pm$} 0.08 & 0.69 {$\pm$} 0.12 & 0.33 {$\pm$} 0.31 & \\
  S2 & 02 56 35.9 & 00 06 27.9 & T & 2.17 {$\pm$} 0.24 & 3.32 {$\pm$} 0.37 & 0.67 {$\pm$} 0.21 & \\
  S3 & 02 56 33.8 & 00 06 28.8 & C & 2.17 {$\pm$} 0.24 & 3.76 {$\pm$} 0.41 & 0.87 {$\pm$} 0.21 & associated with BCG of main component\\
  S4 & 02 56 32.6 & 00 06 30.9 & T & 1.20 {$\pm$} 0.15 & 1.93 {$\pm$} 0.23 & 0.75 {$\pm$} 0.24 & \\
  S5 & 02 56 30.4 & 00 06 01.8 & T & 4.14 {$\pm$} 0.43 & 9.71 {$\pm$} 0.98 & 1.35 {$\pm$} 0.19 & associated with BCG of subcluster\\
  S6 & 02 56 32.2 & 00 05 50.8 & C & 0.42 {$\pm$} 0.08 & 0.59 {$\pm$} 0.12 & 0.54 {$\pm$} 0.38 & foreground source\\
  S7 & 02 56 33.8 & 00 05 02.0 & T & 7.71 {$\pm$} 0.78 & 11.39 {$\pm$} 1.15 & 0.62 {$\pm$} 0.20 & detected in NVSS and FIRST$^c$\\
  \hline
 \end{tabular}
 
 \justify
 $^a$ C: compact; T: resolved with tailed emission.\\
 $^b$ Spectral index between 325 MHz and 610 MHz ($S_\nu \propto \nu^{-\alpha}$). Errors are determined via Monte Carlo methods (see text for details).\\
 $^c$ Extrapolating $S_{610}$ to 1.4 GHz using $\alpha_{_{S7}}$ gives $S_{1400} = 4.61 \pm 0.64$ mJy, which is consistent with the values quoted in Section \ref{sec:arch-Radio}.
 \label{table:radioptsrcs}
\end{table*}

\section{Radio results}
\label{sec:radioresults}
With the angular resolution and short baselines of the GMRT, we are able to investigate emission from both compact sources and extended diffuse structures. In the following, we discuss our results from both the 610 MHz and the 325 MHz datasets.

\subsection{Compact radio sources}
\label{subsec:radioptsrcs}
There are seven bright radio sources in the cluster region identified in both 325 MHz and 610 MHz full-resolution maps, five of which are associated with spectroscopically confirmed cluster members. The 610 MHz HR contours can be seen in the left panel of Figure \ref{fig:GMRT610}, along with source labels. The only source detected in NVSS and FIRST, as discussed in Section \ref{sec:arch-Radio}, is detected in our maps as S7. The flux densities and spectral index we measure for this source, provided in Table \ref{table:radioptsrcs}, imply a consistent 1.4 GHz flux density of 4.61 {$\pm$} 0.64 mJy. 

Several of these sources exhibit resolved tail emission, possibly due to merging activity in the cluster. The BCG of the subcluster is associated with the radio source S5. This source has a wide extension to the west of the galaxy, and although our highest resolution image cannot resolve finer structure within the extended tail, it may be a bent narrow angle tail radio galaxy contorted by ram pressure stripping due to the merger \citep{Bliton.1998.NATs}. The multi-frequency radio properties of all seven sources are given in Table \ref{table:radioptsrcs}. Here and in Section \ref{subsec:alpha}, the spectral indices are determined using a Monte-Carlo simulation, in which we draw from Gaussian flux density distributions with means and widths represented by the flux densities and their uncertainties, respectively. The spectral index and uncertainties are then determined from the median and 68th percentiles of the resulting spectral index distribution.

\begin{figure}
 \includegraphics[width=0.47\textwidth,trim=0 0 0 0,clip]{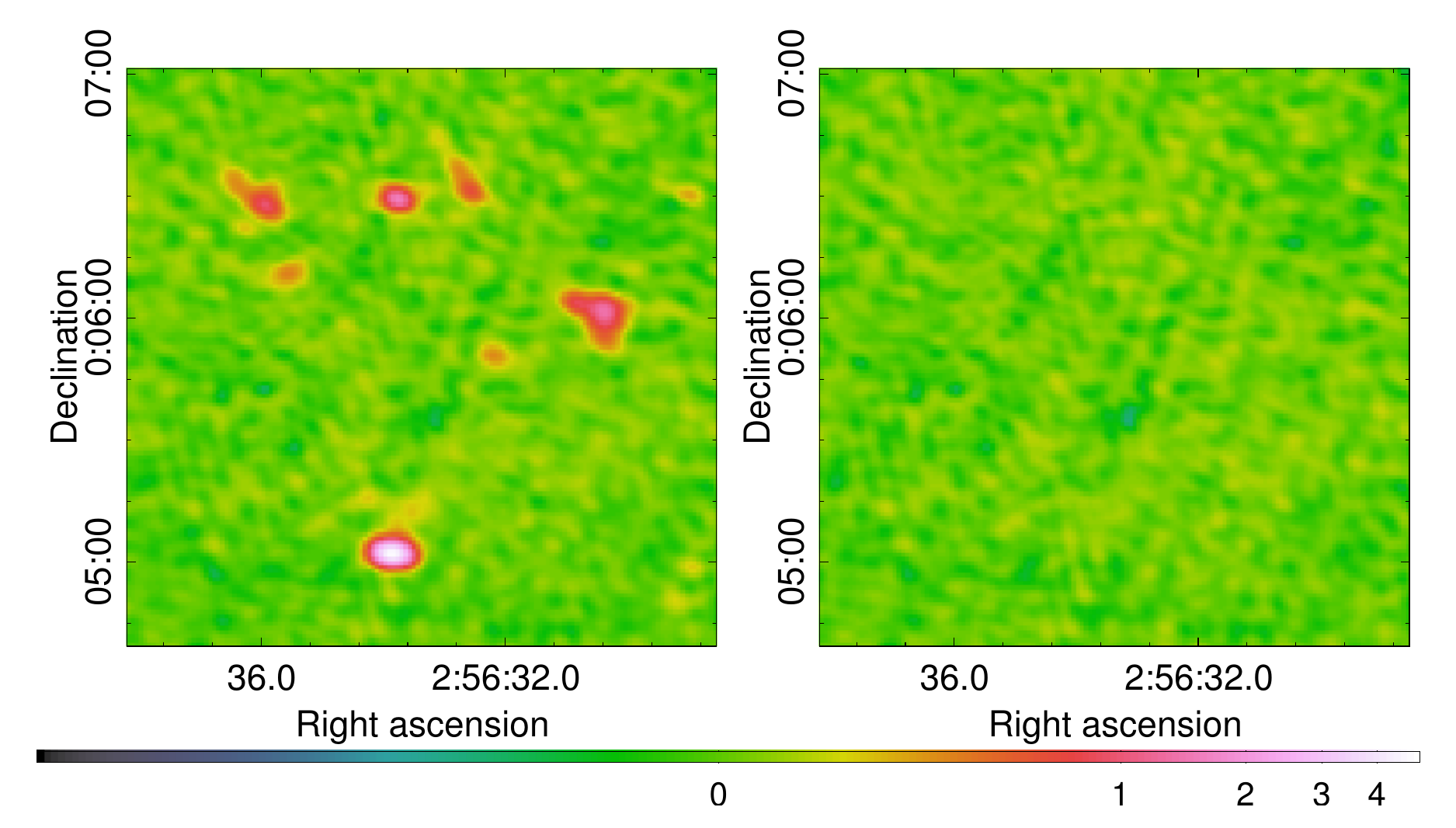}
 \caption{$Left:$ 610 MHz full-resolution (FR) image of the cluster region showing seven radio sources. $Right:$ 610 MHz full resolution image of the same region after subtracting the point source model from the $uv$-data (PSSUB-FR). The colour scale is in units of mJy beam{\per} and is the same for both panels.}
 \label{fig:ptsrcsubB+A}
\end{figure}

\subsection{Point source contamination}
\label{subsec:ptsrccontam}
To unveil any low surface brightness extended cluster emission, the HR radio sources, particularly in the cluster region, have to be removed from the $uv$-data as described in Section \ref{sec:radioobs}. Although the point source removal is reasonably successful, as is clear from the right panel of Figure \ref{fig:ptsrcsubB+A}, it is not exact. In order to quantify the residual (low) level of contamination, we perform a statistical analysis of the LR image using both radio source and random off-source positions in the following way: 
\begin{enumerate}
 \item In the HR image, we select a large number ($>$100) of random off-source positions.
 \item For each position, we calculate the LR map flux density in a LR beam-sized area centred on that position.
 \item From this set of flux densities we calculate the mean, $\mu_{\rm rand}$, and standard deviation, $\sigma_{\rm rand}$, of the distribution. We expect $\mu_{\rm rand}$ to be close to zero for Gaussian noise.
 \item We then select all sources outside of the cluster region that are detected above 5$\sigma$ in the HR map; we find 28 resolved and 53 unresolved sources.
 \item We repeat steps (ii)-(iii), now using the point source positions. $\mu_{\rm ptsrcs}$ quantifies the bias in subtraction of point source emission. $\sigma_{\rm ptsrcs}$ contains both the map uncertainty and a measure of the noise added by the subtraction process, $\sigma_{\rm syst}$, i.e. $\sigma_{\rm ptsrcs}^2 = \sigma_{\rm rand}^2 + \sigma_{\rm syst}^2$.
\end{enumerate}

\begin{figure*}
 \begin{minipage}{\textwidth}
 \centering
   \includegraphics[width=0.48\textwidth,clip]{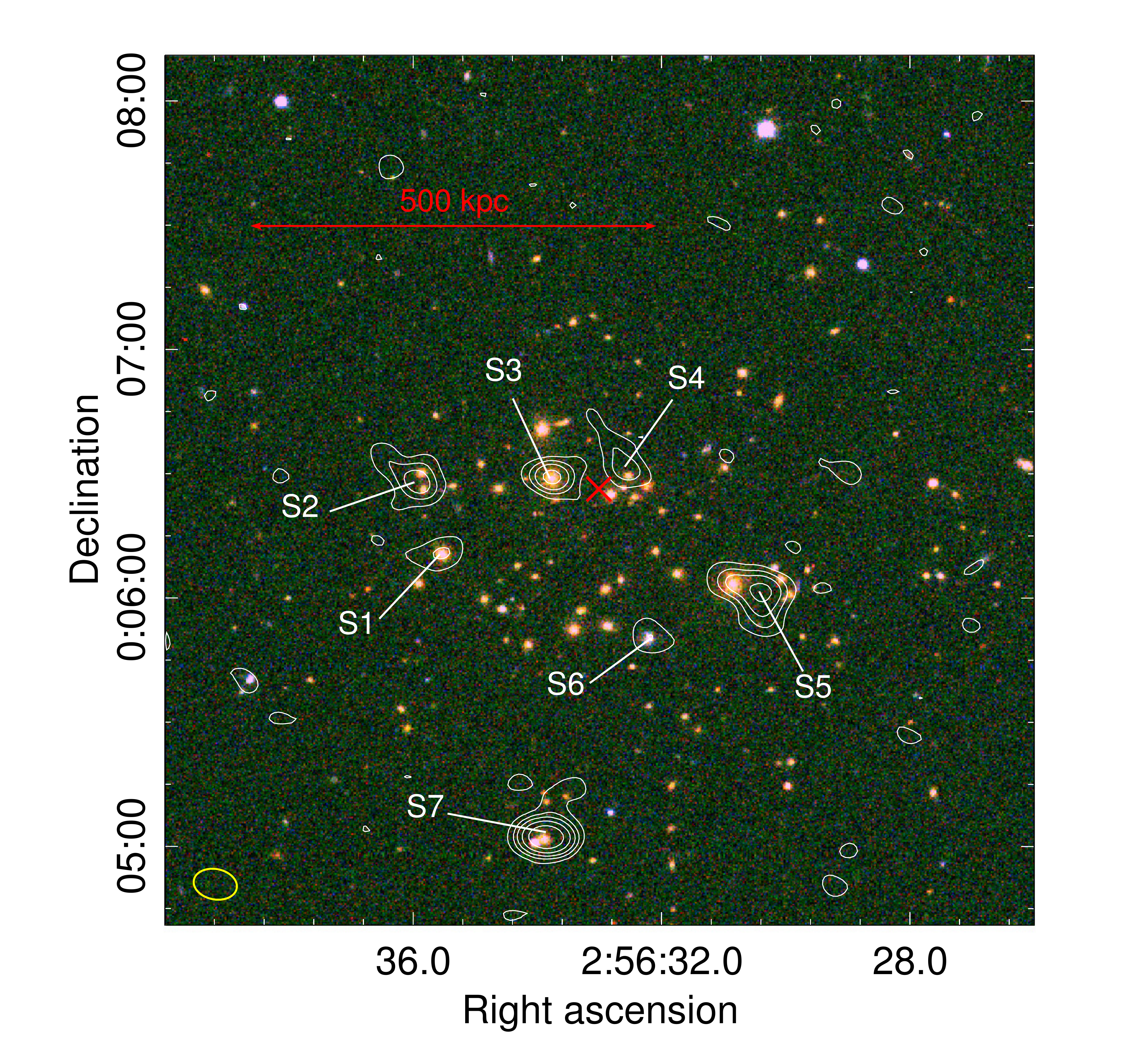}   
   \includegraphics[width=0.51\textwidth,clip,trim=0 0 0 0]{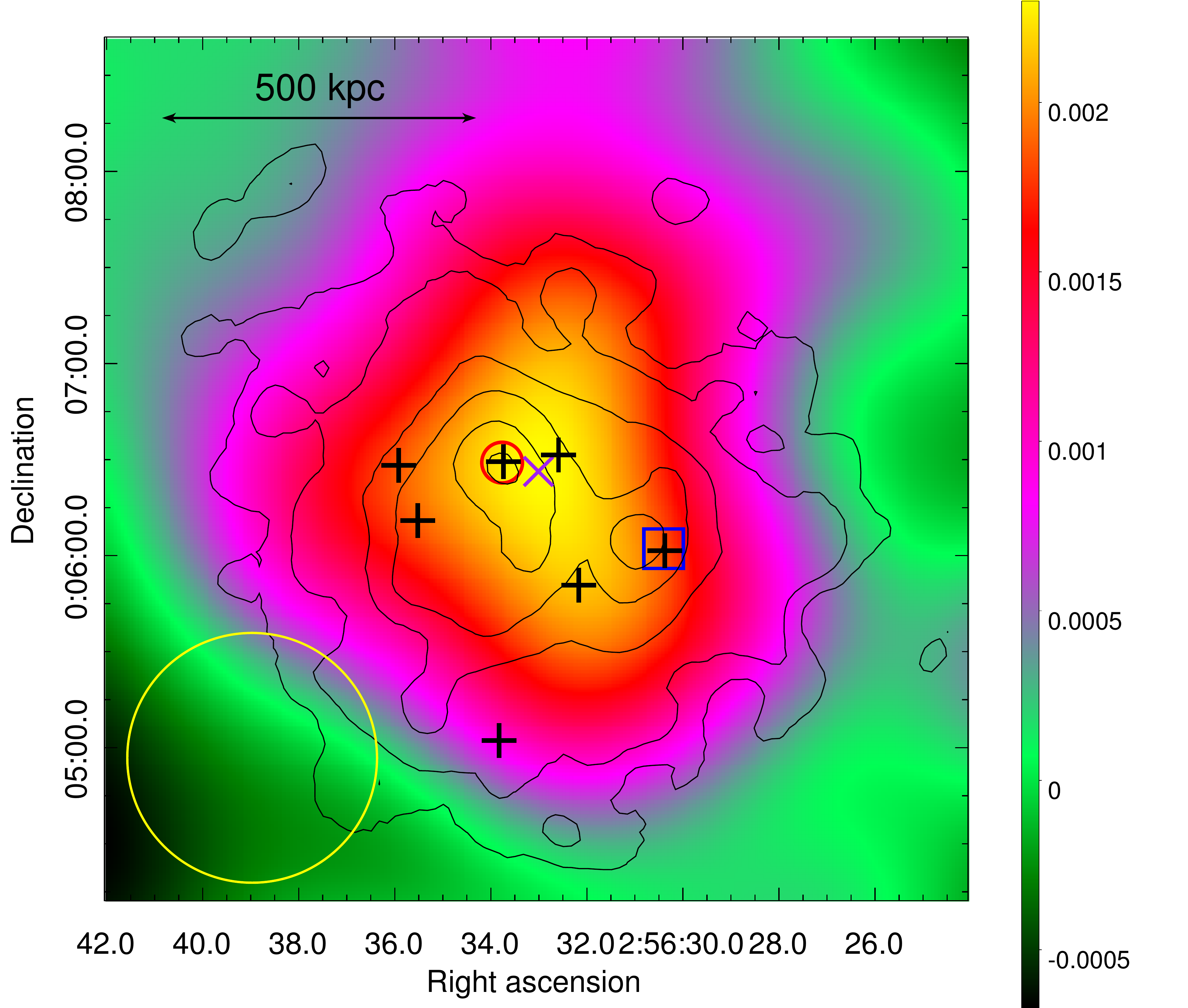}
   \caption{\textit{Left}: GMRT 610 MHz high-resolution (6.5\arcsec $\times$ 5.0\arcsec, p.a. 78.9\degrees) contours of the J0256 emission, overlaid on the SDSS \emph{gri}-band image. The high-resolution (HR) image $1\sigma$ noise level is 31 $\umu$Jy beam\per and the contours are [3,10,20,40,80]$\times 1\sigma$. The HR beam is shown as the yellow ellipse in the lower left corner. Individual radio galaxies are labelled from S1 to S7. Flux densities for these sources can be found in Table \ref{table:radioptsrcs}. The red X marks the position of the SZ peak. \textit{Right}: Smoothed \emph{XMM-Newton} MOS X-ray contours (arbitrary levels from the reprocessed image starting at 12 counts per second per square degree {\textendash} see Figure \ref{fig:xmmepic} in section \ref{sec:xray}), overlaid on the smoothed low-resolution (LR) 610 MHz image of the GRH in J0256. The LR radio image is obtained after subtracting the compact source emission from sources S1 to S7 (positions marked by black crosses). The red circle (
blue square) indicates the BCG of the main (subcluster) component. The positions of the BCGs coincide with the X-ray peaks of each component. The LR synthesised beam (79.6\arcsec $\times$ 76.8\arcsec, p.a. -86.9\degrees) is shown as the yellow ellipse. The purple X marks the position of the SZ peak. The radio colour scale has units of Jy beam\per.}
   \label{fig:GMRT610}
 \end{minipage}
\end{figure*}

The results of this analysis are given in Table \ref{table:ptsrcstats}. We find that we are systematically over-subtracting a low level of point source emission, more so when the sources are resolved. Moreover, the subtraction process does add a small but non-negligible amount of noise into the LR image, as expected. Using the relation in step (v) above, this systematic noise is $\sigma_{\rm syst, 610}$ = 0.3 mJy beam$^{-1}_{_{\rm LR}}$ in the 610 MHz map and $\sigma_{\rm syst, 325}$ = 1.0 mJy beam$^{-1}_{_{\rm LR}}$ in the 325 MHz map. We incorporate these systematic and random residuals into our final flux density measurements (see Section \ref{subsec:fluxes}).

\begin{figure*}
 \begin{minipage}{\textwidth}
  \centering
  \includegraphics[width=0.33\textwidth,clip]{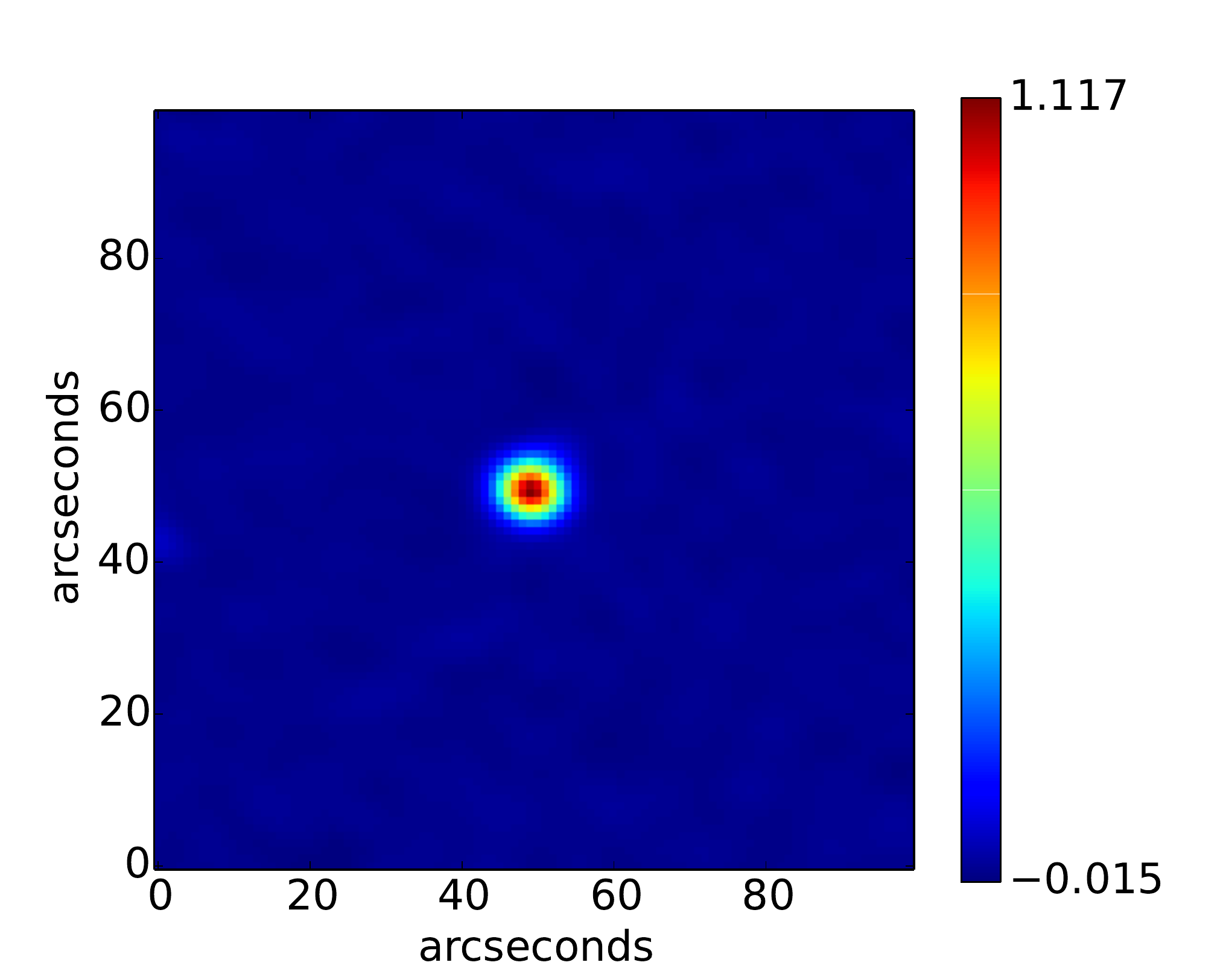}
  \includegraphics[width=0.33\textwidth]{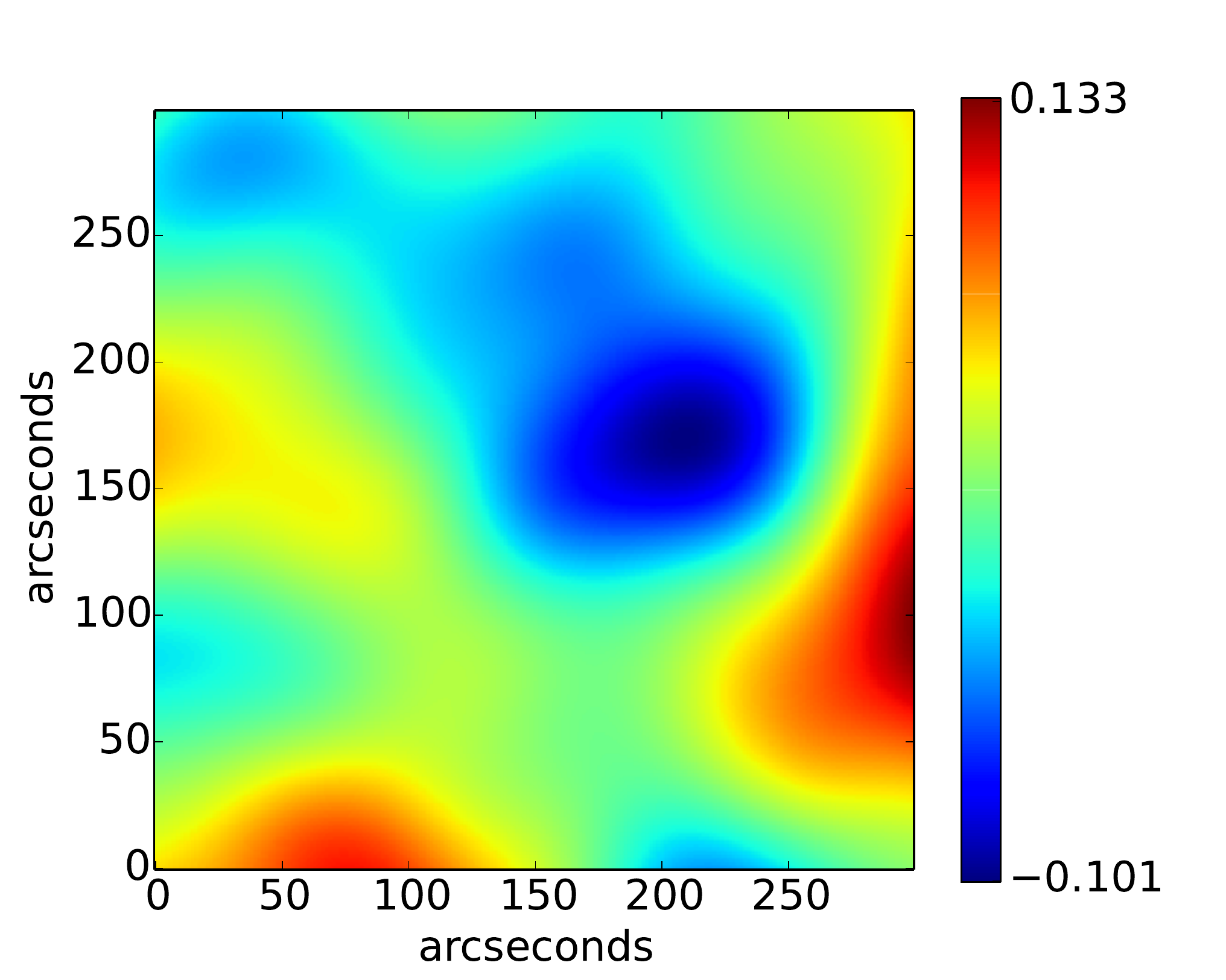}
  \includegraphics[width=0.33\textwidth]{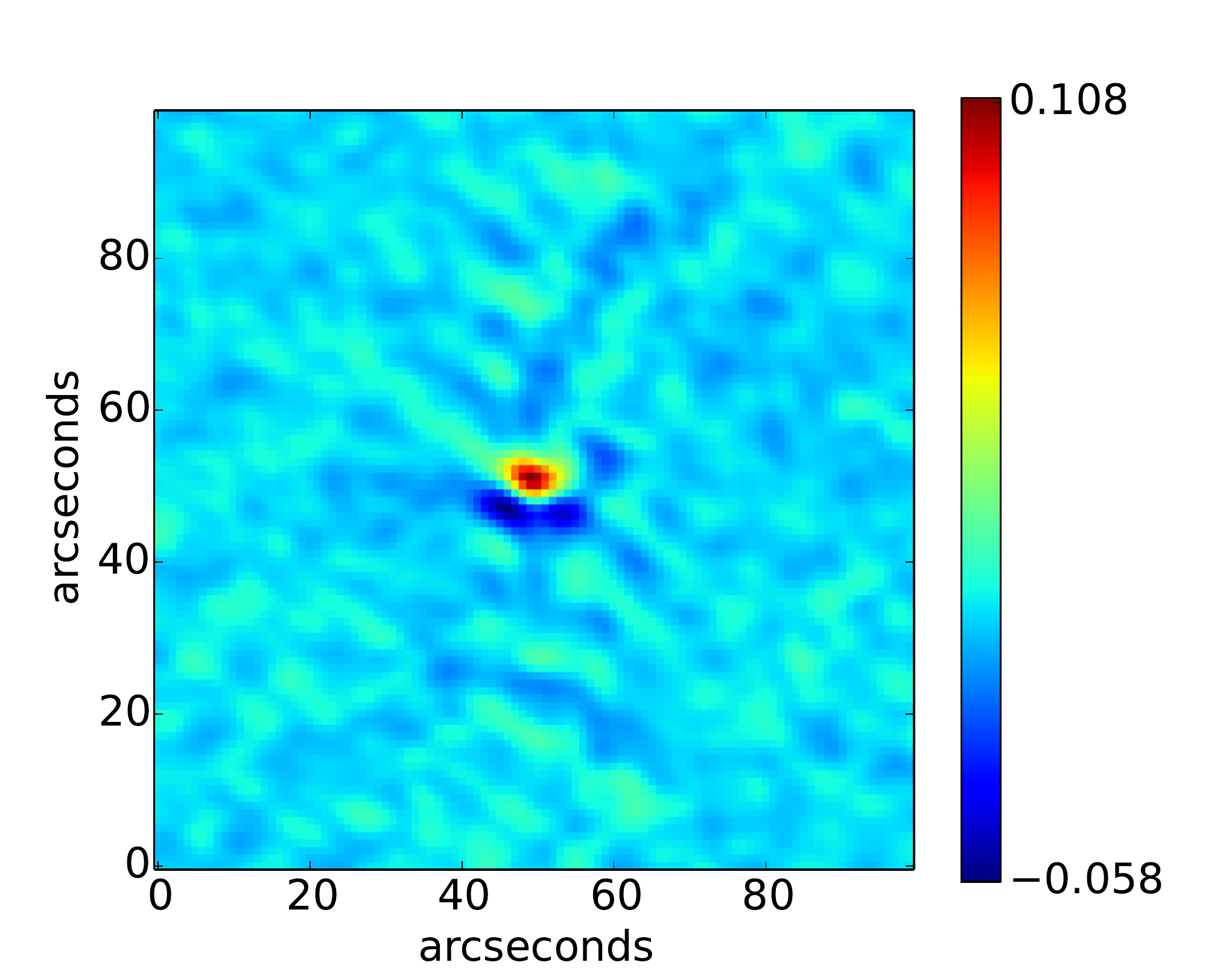}\\
  \includegraphics[width=0.33\textwidth]{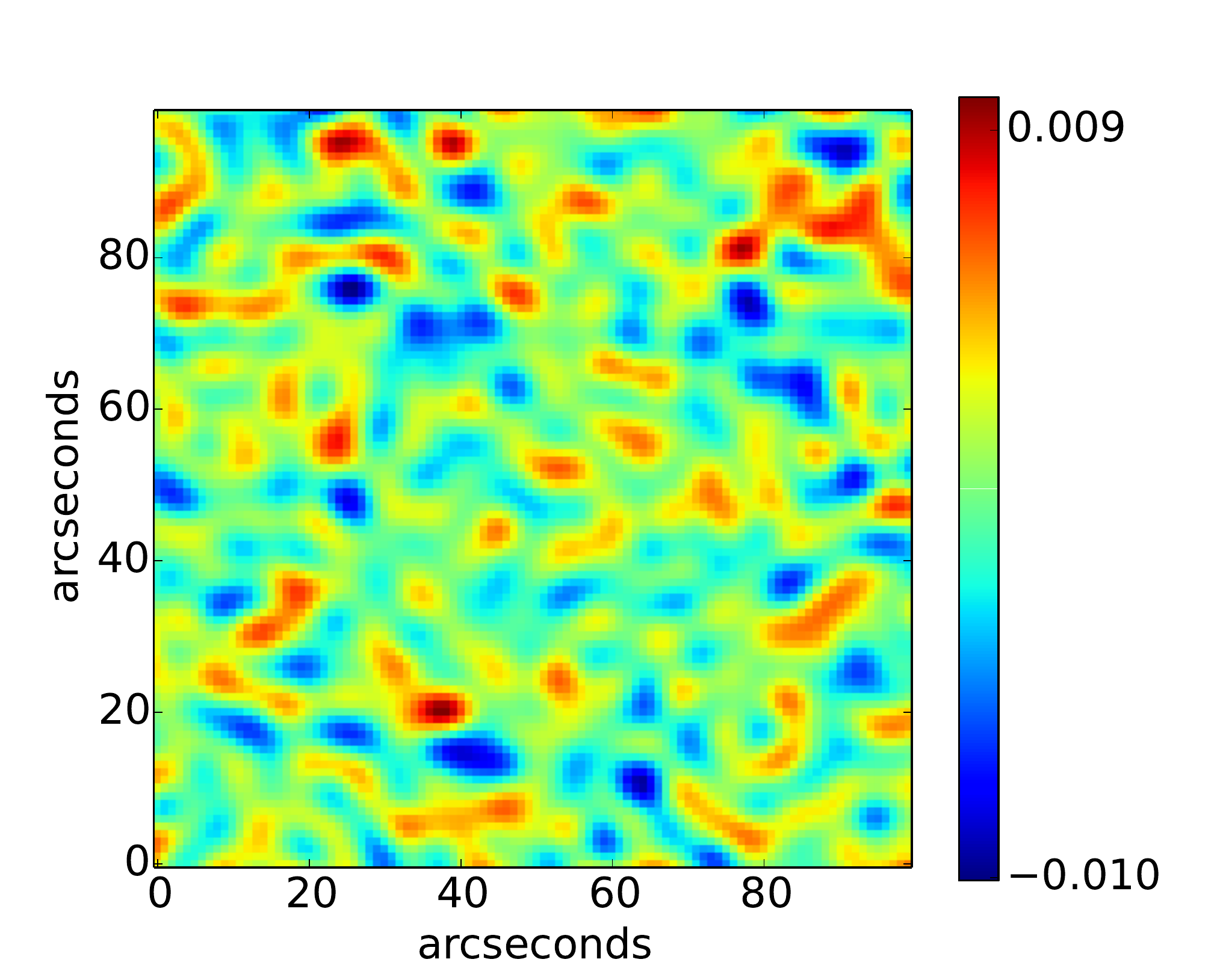}
  \includegraphics[width=0.33\textwidth]{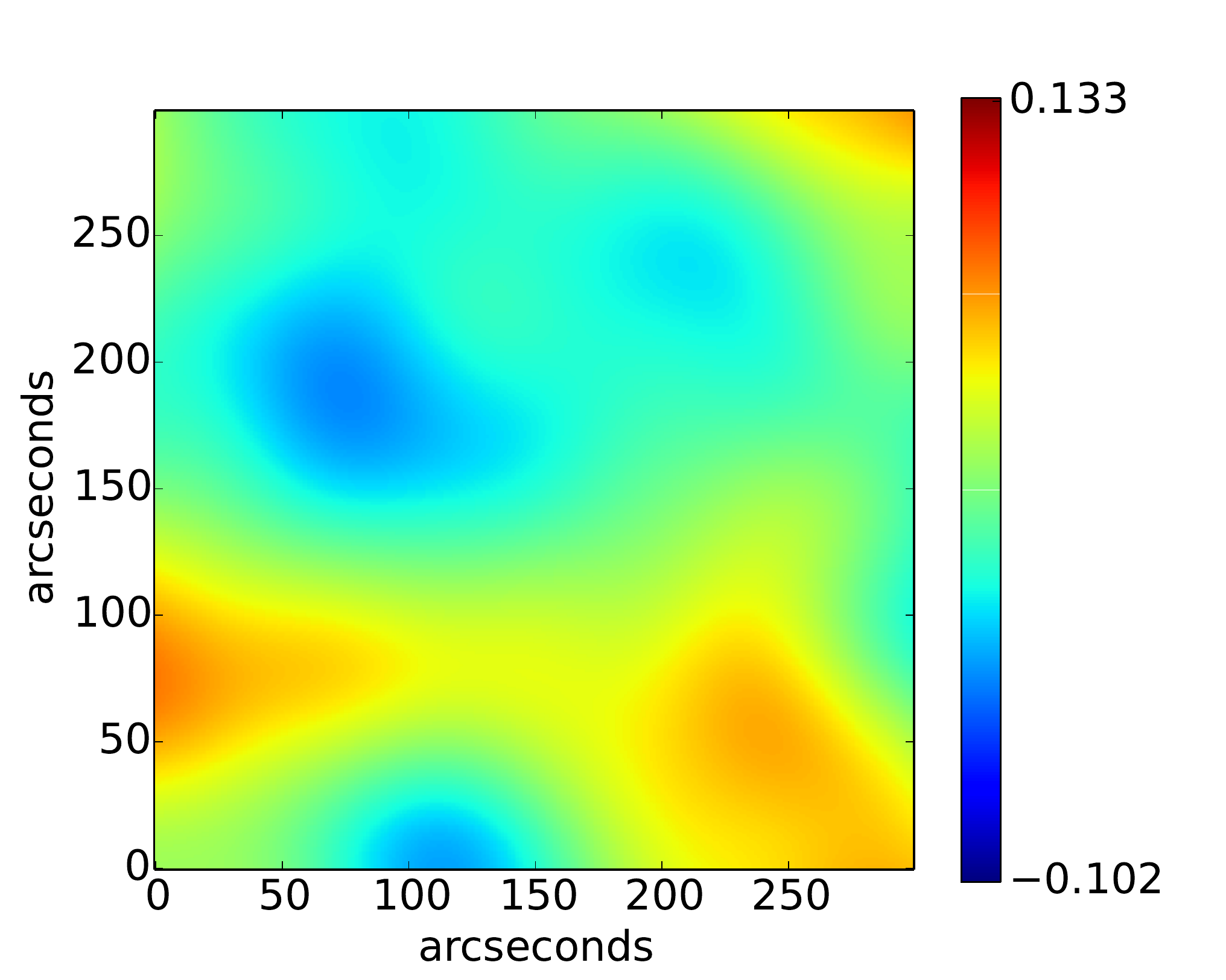}
  \includegraphics[width=0.33\textwidth]{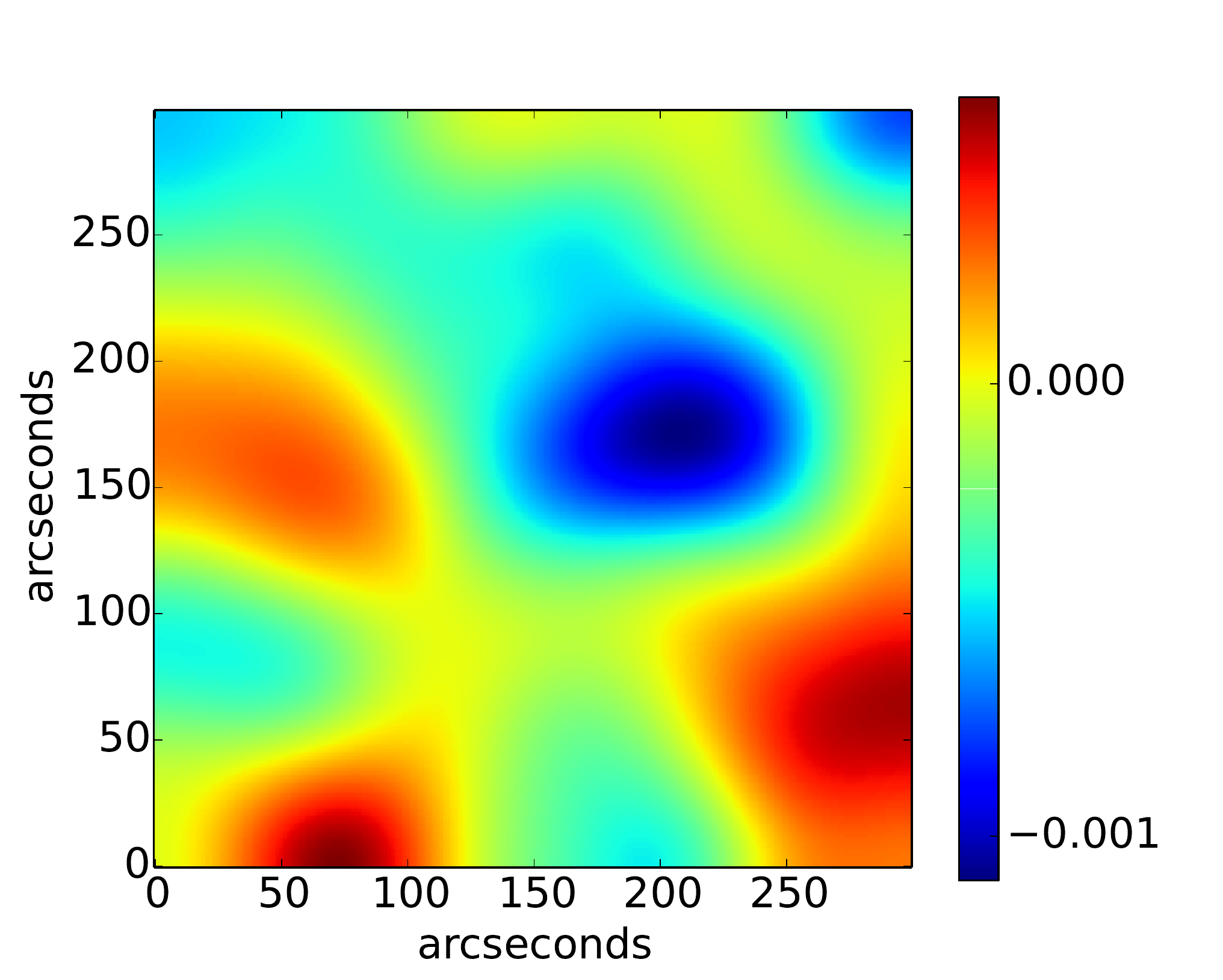}
  \caption{Results from stacking on radio source and random off-source positions in the 610 MHz maps, where all colour scales are in units of mJy beam{\per} and the map axes are in arcseconds. All maps are centred on source positions. {\textit{Left panels:}} Stacked images in the HR map using radio source (upper) and off-source (lower) positions. The elliptical beam is 6{\arcsec} $\times$ 5\arcsec. {\textit{Middle panels:}} Same as the left panels, but for the smoothed, source-subtracted, low-resolution (LR) map. The beam here is 80{\arcsec} $\times$ 70\arcsec. {\textit{Right panels:}} Radio source stacked maps from the PSSUB-FR image (upper) smoothed to the LR beam (lower). }
  \label{fig:stacking}
 \end{minipage}
\end{figure*}

A graphical representation of this process is shown in Figure \ref{fig:stacking}. In the HR and LR maps, we stack on the source and random off-source positions separately. The left panels of Figure \ref{fig:stacking} show the stacked results from the HR map. As expected, the random positions produce a noise-like result and the stacked source positions produce a clear compact source at the centre. 

Repeating this process in the LR image, we find a negative stacked signal slightly off-centre from the source position, in agreement with the over-subtraction implied by in Table \ref{table:ptsrcstats}. The shifted peak is due to the varying noise in the map, shown by the random stacked result (middle panels of Figure \ref{fig:stacking}). We note that the rms of the LR source and off-source stacked maps are comparable. 

As a final check, we stack on the radio source positions in the PSSUB-FR map and smooth this result to the same resolution as the LR map. These results are shown in the right panels of Figure \ref{fig:stacking}. There is a net residual after source subtraction mostly caused by imperfect subtraction of resolved sources, the peak of which is $\sim$10\% of the peak brightness of the average source in the stacked HR map. When we smooth to the same beam as the LR map (lower, right panel), we largely recover the structure of the LR stacked source result (upper, middle panel). 

\begin{table}
 \caption{Results of the systematic and statistical tests to quantify the residual point source contamination in the low-resolution maps. All values are in units of mJy beam$^{-1}_{_{\rm LR}}$.}
 \begin{tabular}{cccccc}
 \hline
  $\nu$ & Quantity & \multicolumn{3}{c}{Source Positions} & Random\\
   (MHz) & & Compact & Resolved & All & Positions\\
  \hline 
  \multicolumn{2}{l}{Number of sources} & 53 & 28 & 81 & 116\\
  \hline
  610 & $\mu$ & -0.075 & -0.082 & -0.077 & 0.013\\
   & $\sigma$ & 0.547 & 0.822 & 0.655 & 0.586\\
   \\
  325 & $\mu$ & -1.073 & -1.920 & -0.971 & 0.273 \\
   & $\sigma$ & 3.109 & 2.470 & 2.693 & 2.503 \\
  \hline 
 \end{tabular}
 \label{table:ptsrcstats}
\end{table}

\subsection{Diffuse emission}
\label{subsec:GRH}
After removal of the radio sources in the field, the LR 610 MHz map, shown in Figure \ref{fig:610LR}, reveals distinct extended emission in the cluster region with a 6$\sigma$ peak above the map noise. The 3$\sigma$ angular extent of the emission is 2.6{\arcmin}, corresponding to a physical scale and largest linear size (LLS) of 0.8 Mpc at the cluster redshift. Due to the centralised position and size of this emission, we classify it as a giant radio halo, making J0256 one of the lowest-mass clusters to host one known to date. The right panel of Figure \ref{fig:GMRT610} shows the 610 MHz GRH overlaid with smoothed X-ray contours. The GRH roughly follows the X-ray emission and is centred on the cluster SZ peak. The GRH radio properties are listed in Table \ref{table:grhstats}. Our LR 325 MHz map is shown in Figure \ref{fig:330LR}. The radio peak of the GRH lies to the west of the cluster SZ peak and is only marginally detected at a significance of 3$\sigma$ in the 325 MHz map.

\begin{table}
 \caption{GRH properties. Subscripts denote frequencies in MHz unless otherwise stated.}
 \label{table:grhstats}
 \centering
  \begin{tabular}[h]{lc}
    \hline
    $S_{610}$ (mJy) & 5.6 {$\pm$} 1.4 \\
    $S_{325}$ (mJy) & 10.3 {$\pm$} 5.3\\
    $\alpha^{610}_{325}$ & 1.0$^{+0.7}_{-0.9}$\\
    $P_{1.4\text{GHz}}$ ($10^{24}$ W Hz\per)$^\dagger$ &  1.0 {$\pm$} 0.3 \\
    LLS$_{610}$ (Mpc)$^\star$ & 0.8 \\
    \hline
  \end{tabular}
  
  \justify
  $^\dagger$ Extrapolated from $S_{610}$ using a spectral index of $\alpha$ = 1.2 {$\pm$} 0.2.\\
  $^\star$ Largest linear size of the GRH, corresponding to 2.6\arcmin.
\end{table}

\subsubsection{Flux measurements}
\label{subsec:fluxes}
The flux density is measured within an aperture of radius 90{\arcsec}, centred on the 610 MHz emission such that all 610 MHz halo flux is captured. From the results of the point source contamination analysis in Section \ref{subsec:ptsrccontam}, the bias at 610 MHz is only at the 1$\sigma$ level, i.e., $\mu_{610,\rm ptsrcs} = -0.077 \pm 0.073 \text{ mJy beam}^{-1}_{_{\rm LR}}$, leading to a 5\% larger corrected flux density for the halo. However at 325 MHz, $\mu_{325,\rm ptsrcs} = -0.971 \pm 0.299 \text{ mJy beam}^{-1}_{_{\rm LR}}$, which is a bias measured at a significance of 3$\sigma$ that leads to a fractional flux density increase of over 50\%. We thus correct the measured flux densities and incorporate the systematic uncertainties introduced by the point source removal into the flux density uncertainties. We also include $\sim$10\% absolute flux calibration and residual amplitude errors \citep{Chandra.2004.GMRTfluxerr}. The final flux density, $S_\nu$, and corresponding uncertainty, $\Delta{S}_{\nu}$, 
are calculated as follows:
\begin{align}
 S_{\nu} = S_{\nu,\rm meas} - \left(\mu_{\nu,\rm ptsrcs}\times N_S\right)\indent \label{eqn:snu}\\
 \Delta{S}_{\nu}^2 = \left(0.1S_{\nu}\right)^2 + \left(\sigma^2_{\rm rms} + \sigma^2_{\rm syst}\right)\times\left(N_S\right) \label{eqn:deltasnu}
\end{align}

\noindent where $\sigma_{\rm rms}$ is the central map noise, $\sigma_{\rm syst}$ is the systematic error due to point source removal, and $N_S$ is the number of independent beams within the flux aperture. We measure integrated halo flux densities of $S_{610}$ = 5.6 {$\pm$} 1.4 mJy and $S_{325}$ = 10.3 {$\pm$} 5.3 mJy. The additional contributions to the flux density uncertainty lower the significance of the 610 MHz detection to 4$\sigma$ which is low, but still reliable. The 325 MHz flux, however, now has a signal-to-noise of less than 2. Higher sensitivity observations at 325 MHz are required to reliably confirm our detection at this frequency. 

\subsubsection{Spectral index}
\label{subsec:alpha}

\begin{figure}
 \centering
 \includegraphics[width=0.48\textwidth]{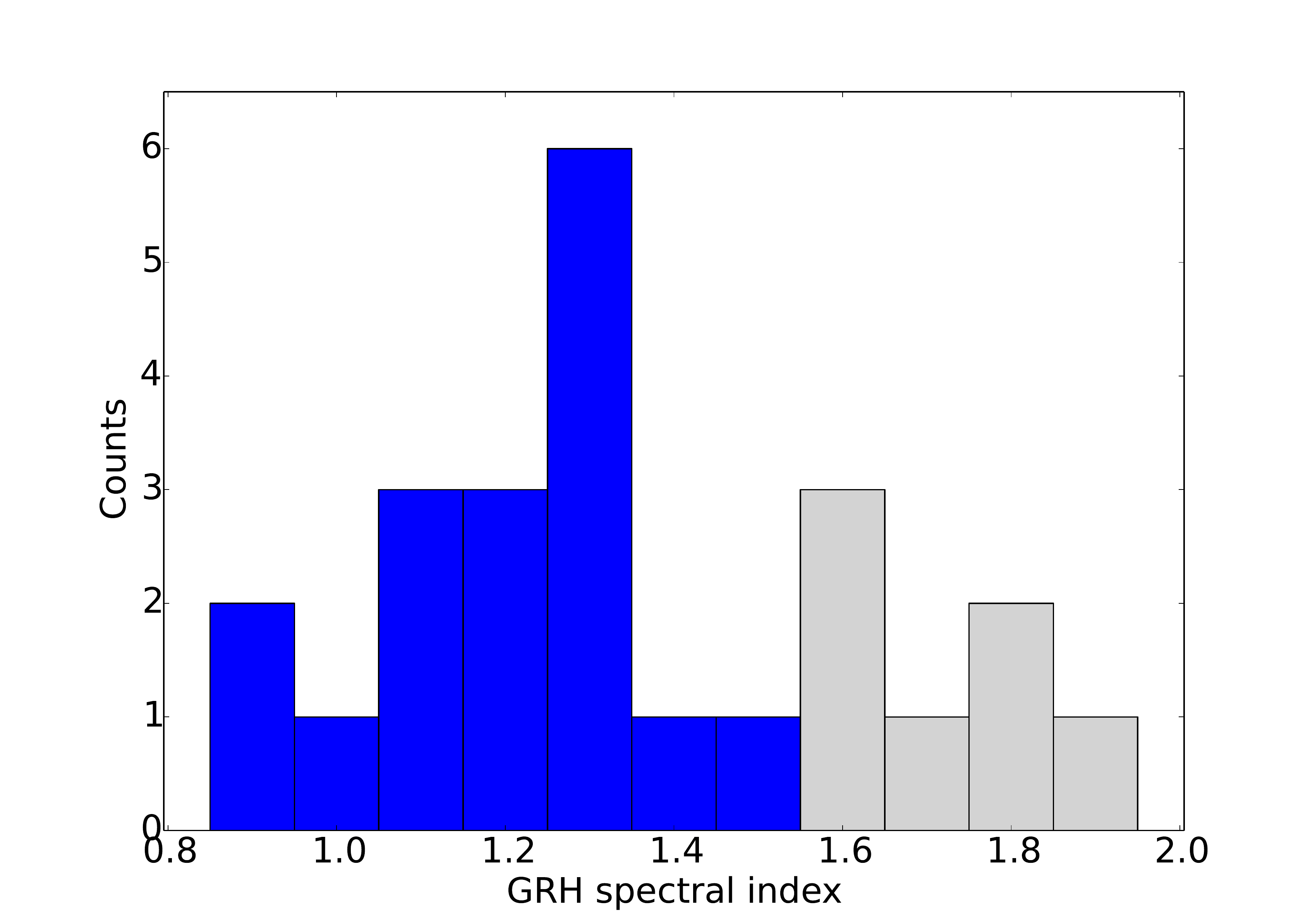}
 \caption{Distribution of all measured GRH spectral indices in the literature. The bulk of the values are taken from \citet{Feretti.2012.Review} with three updated measurements from \citet{Venturi.2013.EGRHS1} and new GRHs from \citet{Bonafede.2014.PLCK287} and \citet{Bonafede.2014.CL1821}. USSRHs ($\alpha \ge 1.6$) are shown in light grey.}
 \label{fig:GRHalphas}
\end{figure}

We can estimate a theoretical spectral index for the GRH in J0256 from the distribution of measured radio halo spectral indices from the literature, shown in Figure \ref{fig:GRHalphas}. Assuming this cluster is in the early stages of merging, based on the X-ray morphology determined by M04 (see Section \ref{sec:arch-Xray} above), we expect J0256 to host a young, and therefore flatter spectrum, radio halo. We therefore exclude the USSRHs ($\alpha \ge 1.6$) from the literature and use the mean and rms of the remaining 17 radio halo spectral indices to determine our theoretical value and error respectively. We determine a spectral index for the typical radio halo population of $\alpha$ = 1.2 {$\pm$} 0.2.

Our measured spectral index, $\alpha^{610}_{325} = 1.0^{+0.7}_{-0.9}$, obtained using $S_{610}$ and the noisy $S_{325}$ measurement, is consistent with the above value. However, given the large uncertainties on $\alpha^{610}_{325}$, driven by the large error on $S_{325}$, we choose to adopt the spectral index of the regular radio halo population, $\alpha$ = 1.2 {$\pm$} 0.2, to extrapolate our measured GRH flux density to other frequencies.

\subsubsection{Radio power}
\label{subsec:radiopower}
The 1.4 GHz GRH radio power, $P_{1.4\text{GHz}}$ is correlated with thermal cluster properties and cluster mass \citep{Cassano.2013.GRHScalRel}. To constrain $P_{1.4\text{GHz}}$, we use our 610 MHz flux density measurement and the assumed spectral index from the previous section to extrapolate a flux density at 1.4 GHz. We account for the effect of redshift on the flux density and apply a k-correction to calculate a halo radio power of $P_{1.4\text{GHz}}$ = (1.0 {$\pm$} 0.3) $\times 10^{24}$ W Hz{\per} in the cluster rest frame. The error on $P_{1.4\text{GHz}}$ is propagated from the spectral index uncertainties. We note that the radio power is consistent with the non-detections in NVSS, FIRST, and VLSS, as it corresponds to a GRH surface brightness far below the noise levels of these surveys.

\begin{figure}
 \centering
   \includegraphics[width=0.39\textwidth,trim=50 0 50 0]{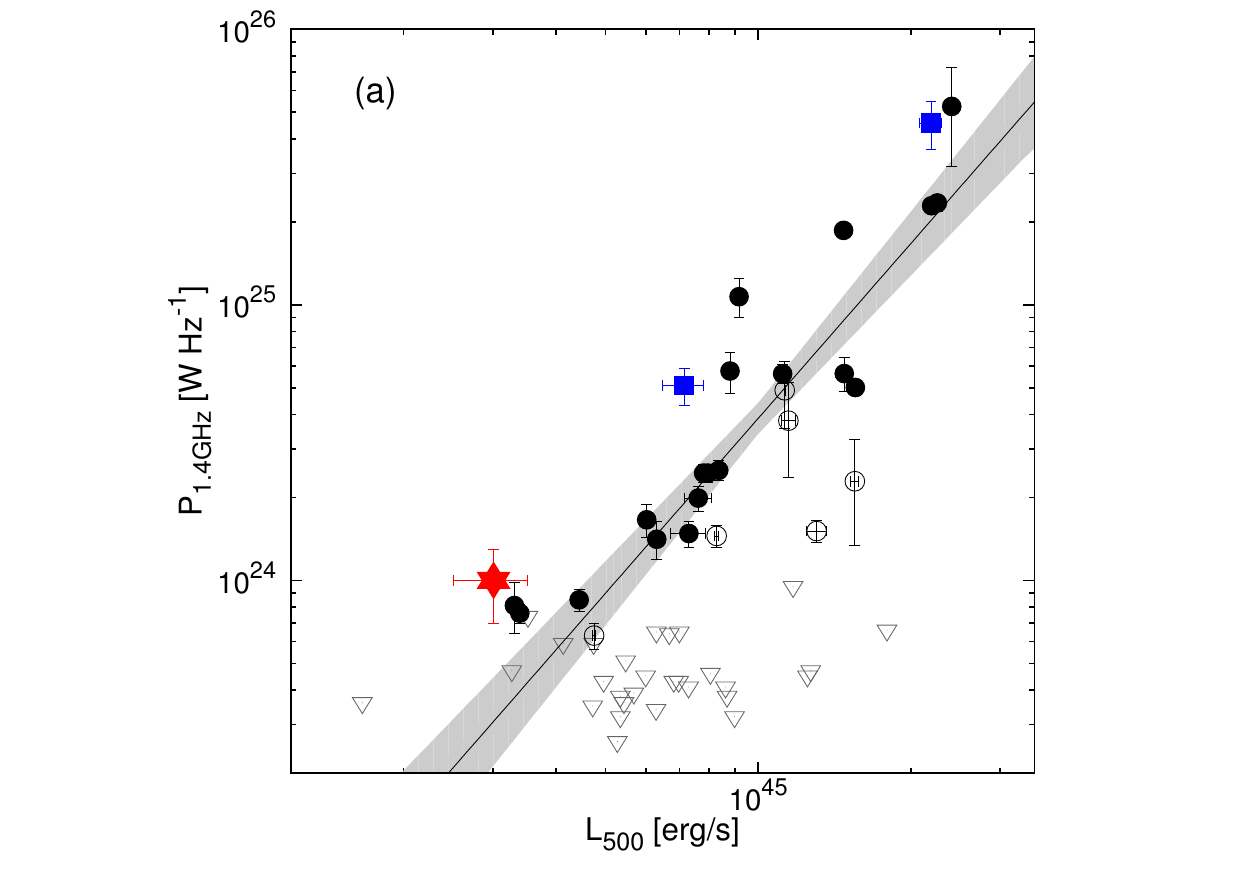}\\
   \includegraphics[width=0.39\textwidth,trim=50 0 50 0]{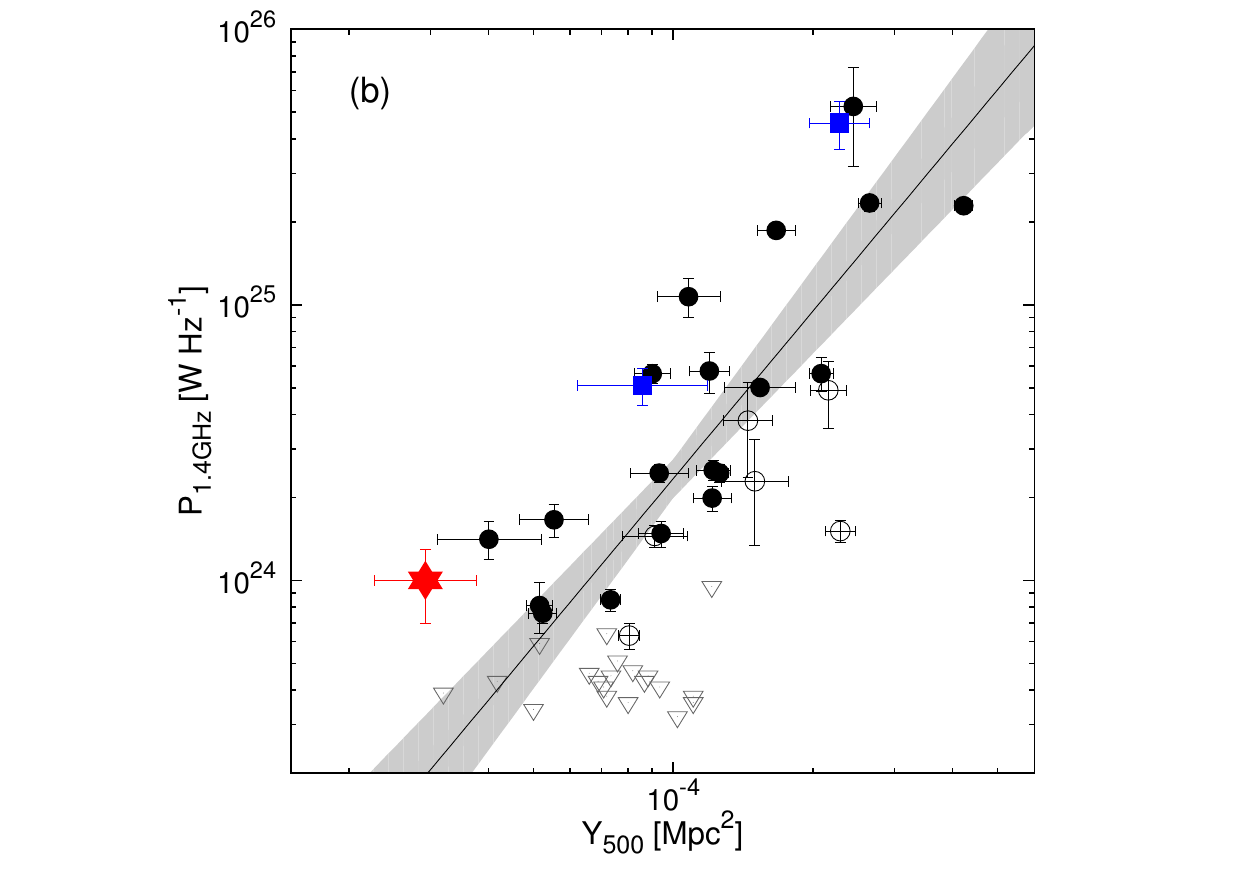}\\
   \includegraphics[width=0.39\textwidth,trim=50 0 50 0]{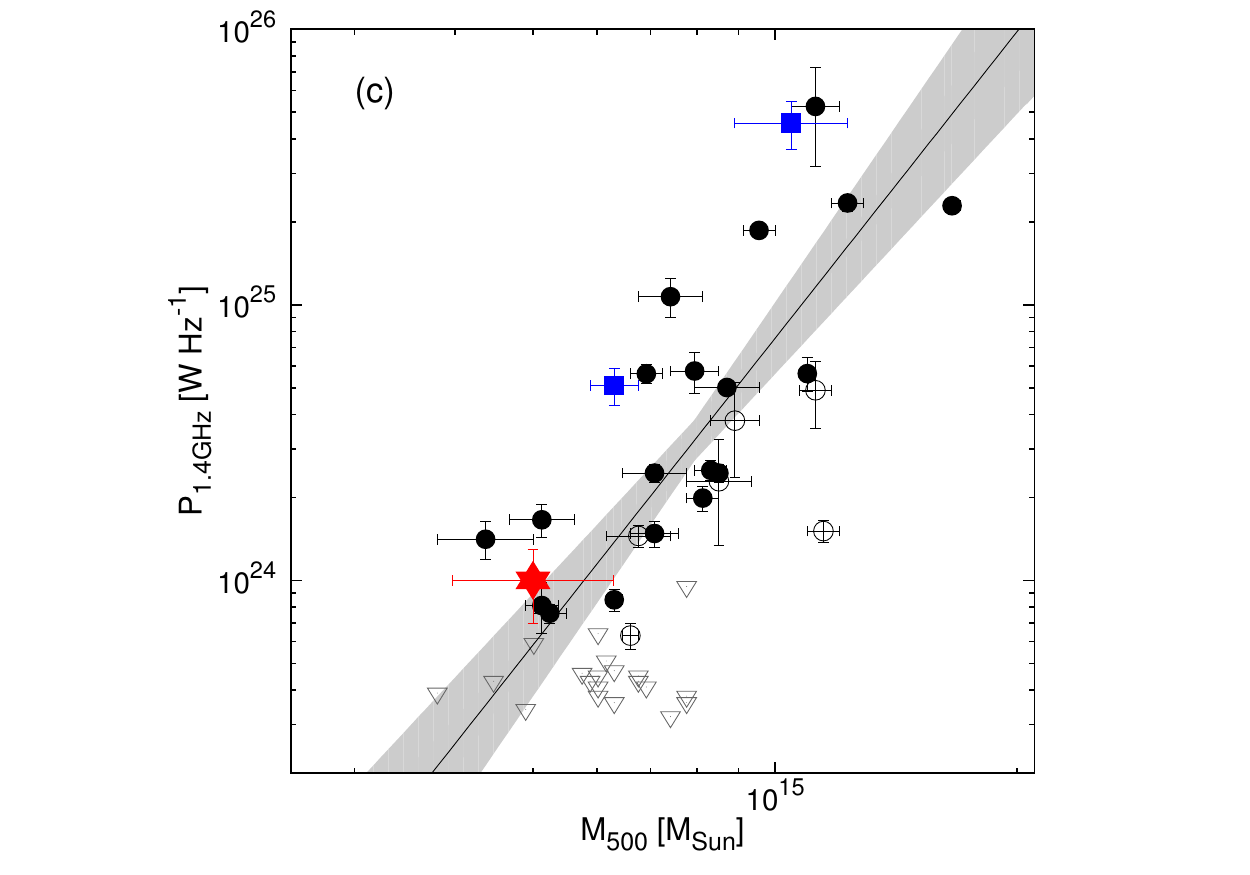}
   \caption{Radio halo detections and upper limits from the literature showing correlations between the 1.4 GHz radio power and cluster thermal parameters {\textemdash} (a) $P_{1.4}$ vs $L_{\rm X}$, (b) $P_{1.4}$ vs $Y_{500}$, and (c) $P_{1.4}$ vs $M_{\rm 500,SZ}$. Black solid (open) circles and grey open triangles are giant radio halos (USSRHs) and upper limits, respectively, from \citet{Cassano.2013.GRHScalRel}, with recent GRHs in PLCK147.3-16.6 \citep{vanWeeren.2014.PLCK147} and El Gordo \citep{Lindner.2014.ElGordo} shown as blue squares. The position of J0256 is shown as a red star. The best fit to the GRH detections and associated 95\% confidence interval is from \citet{Cassano.2013.GRHScalRel} and are shown by the black line and grey shaded region, respectively.}
   \label{fig:P1.4corrs}
\end{figure}

J0256 is shown as the red star on the radio power correlations in Figure \ref{fig:P1.4corrs}. The cluster lies within the scatter, and on the same side, of all three correlations from the literature. J0256 appears to lie slightly further away from the $P_{1.4\text{GHz}}$--$Y_{500}$ relation, compared to its relative position in the other planes. However, the position of the cluster, relative to the distance away from each correlation, is consistent within the error bars for the cluster mass and thermal parameters.

\begin{figure}
 \centering
 \includegraphics[width=0.48\textwidth, clip=true, trim=0 0 50 0]{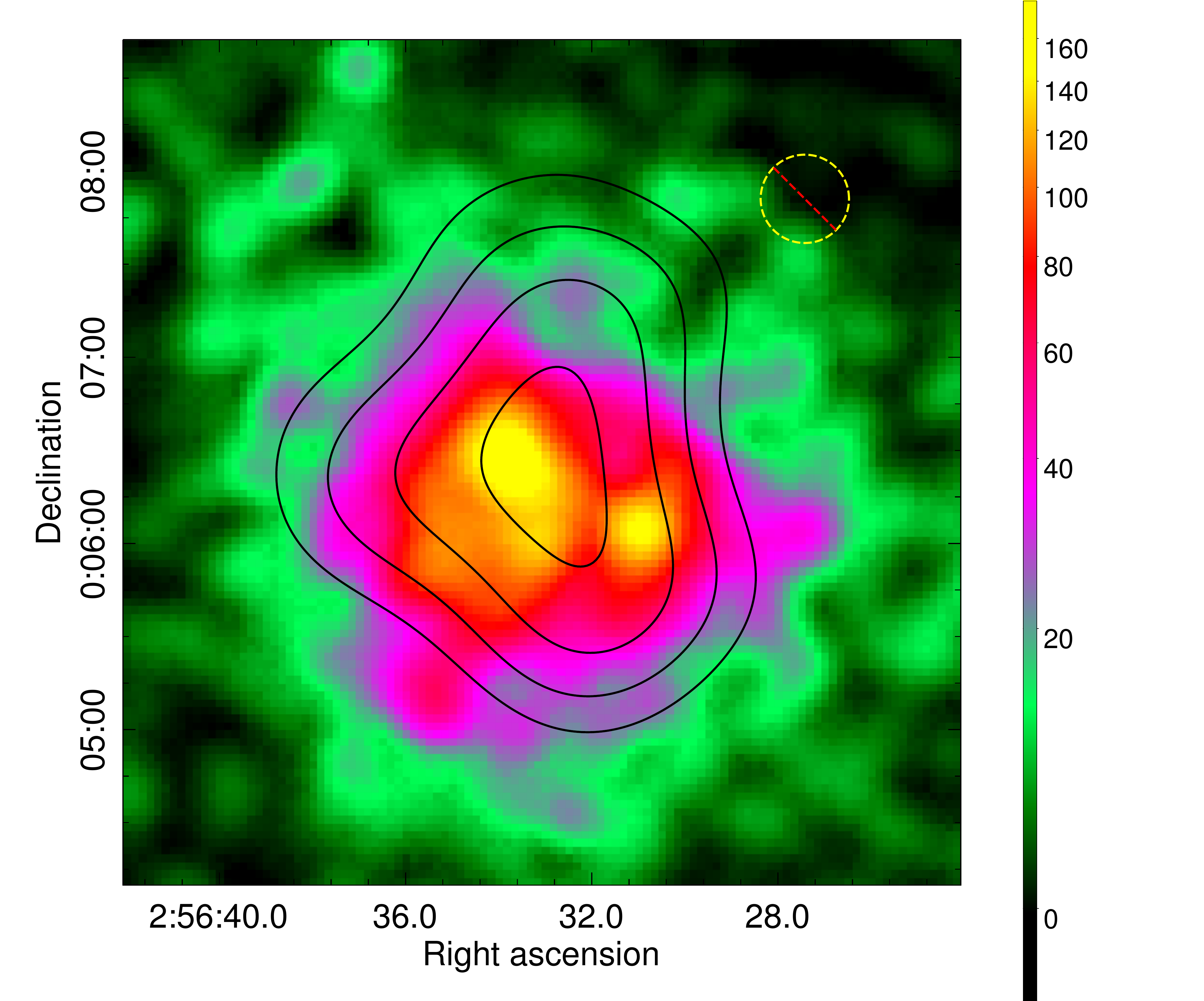}
 \caption[Combined MOS \textit{XMM-Newton} image of J0256.]{Combined 0.5-2.0 keV MOS1+MOS2 \textit{XMM-Newton} image of ACT-CL J0256.5+0006 with the 610 MHz radio halo contours overlaid. The contours start at 3$\sigma$ = 0.001 Jy beam\per, and increase in steps of 1$\sigma$. The X-ray image has been exposure-corrected and background-subtracted. The region masked after removing point sources is indicated by the yellow, dashed, excluded circle. The image is unbinned and has been smoothed by a Gaussian with a kernel radius of 6 pixels (1 px = 2.5\arcsec). The units of the colour scale are counts per second per square degree.}
 \label{fig:xmmepic}
\end{figure}

\section{Cluster morphology}
As current observations favour a theory of merger-driven radio halo formation, it is important to understand the dynamical state of J0256. With the X-ray and optical redshift information available to us, we can perform a morphological analysis of J0256.

\subsection{X-ray morphology}
\label{sec:xray}
Visual inspection of the reprocessed J0256 X-ray image in Figure \ref{fig:xmmepic} shows the cluster to be disturbed.  This image is produced by following the ESAS reduction thread for extended X-ray sources\footnote{\url{http://heasarc.gsfc.nasa.gov/docs/xmm/esas/esasimage/esasimage_thread-str.html}}, in which the Al and Si lines are modelled in \textsc{xspec}. The image has been both exposure-corrected and background-subtracted. We exclude the pn camera data as the pn CCD is marginally flared and has a chip gap near the cluster core. Point sources were removed during the reprocessing using the \textsc{cheese} task in the standard \textsc{XMM SAS} tools\footnote{\url{http://xmmssc-www.star.le.ac.uk/SAS/xmmsas_20121219_1645/doc/cheese/index.html}}. The masked regions are shown in Figure \ref{fig:xmmepic} by yellow, dashed, excluded circles.

In order to quantify the level of substructure in the reprocessed \emph{XMM-Newton} combined MOS1 + MOS2 image, we follow the work of \citet{Cassano.2010.GRHMergerConn} by calculating three morphological parameters. To determine the measurement uncertainty on each of our parameters, we adopt the simulation method of \citet{Bohringer.2010.MorphParams} whereby a Poisson resampled X-ray image is used to compute the standard deviation of a parameter measurement, which is then used to estimate the measurement uncertainty. 

\subsubsection{Concentration parameter, $c_{SB}$}
\label{subsec:concentration}
The concentration parameter, proposed by \citet{Santos.2008.MorphParams} as a probe of cluster substructure, is the ratio of the cluster core and the larger-scale X-ray surface brightnesses. We calculate the concentration parameter as
\begin{equation}
 c_{SB} = \frac{S (< 100\text{ kpc})}{S (< 500\text{ kpc})},
\end{equation}

\noindent where $S$ is the X-ray surface brightness within a particular radius, centred on the X-ray peak. We determine a value of $c_{SB} = 0.151 \pm 0.007$ for J0256.

\subsubsection{Centroid shift, $w$}
\label{subsec:centroidshift}
\citet{Poole.2006.MergerImpact} show that, compared to other X-ray morphological estimators, the centroid shift is the most sensitive to cluster dynamical state and least sensitive to cluster image noise. It is defined as the rms deviation of the projected separation between the X-ray peak and the centre of mass in units of the aperture radius, $R_{ap}$, computed in a series of concentric circular apertures centred on the cluster X-ray peak \citep{Mohr.1993.XSubstruct, OHara.2006.XMorphParams, Maughan.2008.MorphParams}. Following \citet{Cassano.2010.GRHMergerConn}, the aperture radius is decreased in steps of 5\% from a maximum aperture of radius $R_{ap} = 500$ kpc to 0.05 $R_{ap}$. We compute the centroid shift as
\begin{equation}
 w = \left[\frac{1}{N-1} \sum_i\left(\Delta_i - \langle\Delta\rangle\right)^2\right]^{1/2} \times \frac{1}{R_{ap}},
\end{equation}

\noindent where $\Delta_i$ is the distance between the X-ray peak and the centroid of the $i$th aperture. We measure a value of $w = 0.045 \pm 0.006$ for J0256.

\subsubsection{Power ratio, $P_3/P_0$}
\label{subsec:powerratio}
The power ratio of a cluster is calculated using a multipole decomposition of the potential of the two-dimensional projected mass distribution. The idea of using the power ratio of the X-ray surface brightness to probe the underlying mass distribution was first introduced by \citet{BuoteTsai.1995.MorphParams} and has since been widely used as an indication of substructure within a cluster \citep{Jeltema.2005.MorphParams, Ventimiglia.2008.MorphParams, Bohringer.2010.MorphParams, Cassano.2010.GRHMergerConn}. We use the normalised hexapole moment, $P_3/P_0$, which is the lowest power ratio moment providing a clear measure of substructure \citep{Bohringer.2010.MorphParams}, calculated in an aperture of radius $R_{ap} = 500$ kpc centred on the X-ray cluster centroid. For J0256, we calculate a value of $P_3/P_0 = (3.76 \pm 0.30) \times 10^{-6}$.

\subsubsection{Comparison with the literature}
Using the methods described in Sections \ref{subsec:concentration}-\ref{subsec:powerratio}, \citet{Cassano.2010.GRHMergerConn} study the morphological parameters for all clusters in the GMRT Radio Halo Survey \citep{Venturi.2007.GRHS1, Venturi.2008.GRHS2} and find a link between cluster dynamical state and the presence of a radio halo. They define a cluster to be dynamically disturbed if its morphological parameters satisfy the following conditions: $c_{SB} <  0.2$, $w > 0.012$ and $P_3/P_0 > 1.2 \times 10^{-7}$. The majority of dynamically disturbed clusters are found to show radio halo emission. All of the parameter values we determine in our analysis of J0256 ($c_{SB} = 0.151 \pm 0.007$, $w = 0.045 \pm 0.006$ and $P_3/P_0 = (3.76 \pm 0.30) \times 10^{-6}$) satisfy the above conditions for a merging cluster, as expected. 

We note that the \citet{Cassano.2010.GRHMergerConn} results were obtained using \textit{Chandra} data whereas our results are obtained with \textit{XMM-Newton} data, which has a larger PSF. To investigate the effect of the different instruments on the various morphological parameters, we use archival \textit{Chandra} and \textit{XMM-Newton} data on a known merging cluster, A2631, and compare the derived morphological parameters from each image. The exposure times for the Chandra and XMM observations, after flare rejections, are 16.8 ks and 13.4 ks, respectively. We find that the resolution difference between the two instruments has a negligible effect on the concentration or centroid shift parameters. However, the power ratio is higher in the XMM image by a factor of between 2 and 5, depending on the level of smoothing applied. Even with a reduction by a factor of five, the $P_3/P_0$ value for J0256 (7.5 $\times 10^{-6}$) is still well above the threshold of $1.2 \times 10^{-7}$ for dynamically disturbed 
clusters.

\subsection{Optical redshift distribution}
\label{sec:optical}
\begin{figure}
 \centering
 \includegraphics[width=0.5\textwidth,clip,trim=50 0 50 0]{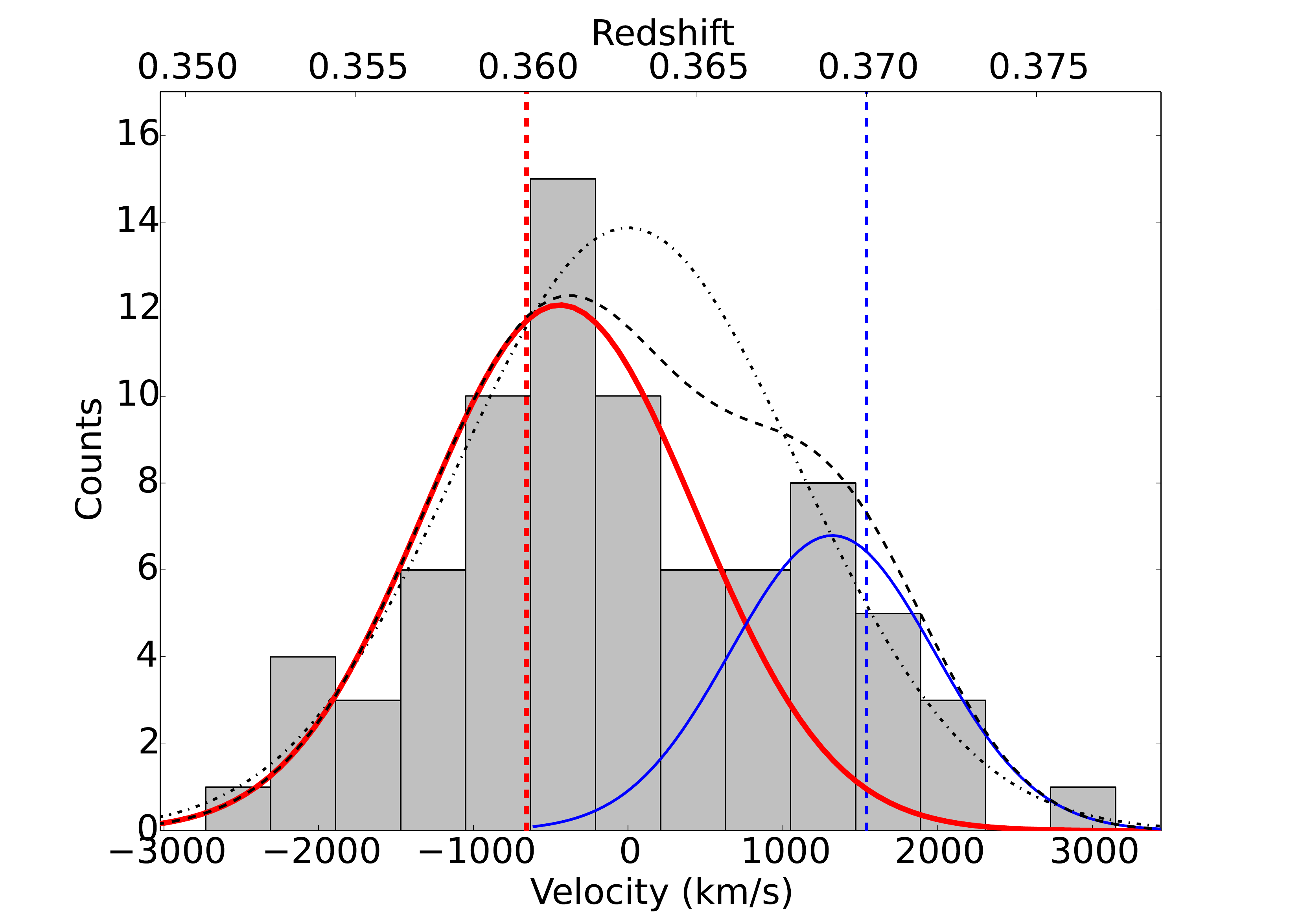}
 \caption{Histogram showing the redshift distribution for 78 spectroscopically confirmed cluster members. Here $v = 0$ is defined as the cluster systemic redshift of $z = 0.363$, and the bin width is 420 km s\per. A bimodal fit of two Gaussians (dashed black) is shown along with the constituent main component (thick red; $\mu = 0.361 \pm 0.001, \sigma = 0.004 \pm 0.001$) and sub-component (thin blue; $\mu = 0.369 \pm 0.002, \sigma = 0.003 \pm 0.001$). The vertical thick red (thin blue) dashed line shows the velocity of the BCG for the main (subcluster) component. A single Gaussian fit ($\mu = 0.363 \pm 0.002, \sigma = 0.005 \pm 0.001$) is shown by the dot-dashed black curve. }
 \label{fig:zvhisto}
\end{figure}

X-ray morphological parameters are largely insensitive to substructure along the line of sight. To gauge any disturbed morphology in this direction, we use the redshift distribution of 78 spectroscopically confirmed cluster member galaxies (see Section \ref{sec:arch-Opt} above). This distribution is shown in Figure \ref{fig:zvhisto}; there is an indication of bimodal structure in the histogram.

\begin{table*}
 \centering
 \caption{GMM statistics from the redshift distribution of 78 cluster members. All errors are at the 1$\sigma$ level.}
  \begin{tabular}[h]{lccccccc}
    \multicolumn{1}{l}{kurtosis, $K$} & -0.260 \\
    \multicolumn{1}{l}{peak separation, $D$} & 2.64 $\pm$ 0.82 \\
    \hline\hline
    Distribution type & \multicolumn{4}{c}{Statistics} & \multicolumn{3}{c}{Bootstrapping (\%)$^\ddagger$} \\
      & $n$ & $\mu$ & $\sigma^2$ & $\log{L} ^\dagger$ & $K$ & $D$ & $\chi^2$ \\
    \hline
    Unimodal  & 78 & 0.363 $\pm$ 0.001 & 0.005 $\pm$ 0.000 & 299.6 & - & - & - \\
    \hline
    Bimodal, multi-variance & 53.9 $\pm$ 15.9 & 0.360 $\pm$ 0.002 & 0.004 $\pm$ 0.001 & 300.7 & 49.0 & 46.6 & 69.4\\
     & 24.1 $\pm$ 15.9 & 0.369 $\pm$ 0.003 & 0.003 $\pm$ 0.001 & \\
    \hline
  \end{tabular}
  \label{table:GMM}
  
  \justify
  $^\dagger$ The maximum log likelihood to which the fit converges. The difference in $\log{L}$ values defines a $\chi^2$ proxy.\\
  $^\ddagger$ Measure of how likely it is that the same statistic can be drawn from a unimodal model.
  
\end{table*}

\subsubsection{Statistical analysis using GMM}
\label{sec:GMM}
To gauge its significance, we perform a Gaussian mixture model (GMM) analysis of the member galaxy redshifts. We use the GMM code developed by \citet{MuratovGnedin.2010.GMM} to fit a 2-mode Gaussian mixture to our data and compare it to a unimodal fit. The code calculates the kurtosis of the distribution, $K$, and the maximum log likelihood, $\log{L}$, to which each model converges. For a bimodal fit, the peak separation of the modes relative to their widths, $D$, is also calculated. A statistically significant bimodality would have $K < 0$, $D > 2$, and a log-likelihood value greater than that for a unimodal fit. Parametric bootstrapping of the unimodal distribution is performed to determine the probabilities of the observed $K$, $D$, and $\log{L}$ difference values being sampled from a unimodal distribution. The latter probability defines the confidence interval at which a unimodal fit can be rejected.

The results of our analysis are given in Table \ref{table:GMM}. The multi-variance bimodal mixture model and unimodal Gaussian fits are superimposed on the distribution in Figure \ref{fig:zvhisto}, shown by the dashed and dot-dashed curves respectively. The data satisfy the $K < 0 $ and $D > 2$ criteria for bimodality, with the largest $\log{L}$ value coming from the multi-variance bimodal fit. The improvement in the $\log{L}$ value for the multi-variance bimodal model relative to the unimodal model is not significant due to the difference in degrees of freedom; a likelihood-ratio test indicates that the bimodal fit is rejected in favour of the unimodal fit at 53\%. According to the parametric bootstrapping, the unimodal distribution is consistent with the data at the 69\% level when only the $\log(L)$ probability is considered, with bootstrapped probabilities of $K$ = 49\% for the kurtosis, and $D$ = 47\% for the peak separation. A unimodal fit thus cannot be ruled out.

However, statistical tests run on mock bimodal datasets, with the same population ratio and number of members as our real data, show that the GMM test results in a $\log(L)$ probability of 70\% or higher about 10\% of the time. An unambiguous bimodal preference is only consistently achieved once the total population size is greater than 200. This implies that, when the distribution size is small, the GMM test could show a slight preference for a unimodal fit even when the input redshift distribution is bimodal, given the population ratio of our true sample. Therefore, with the available number of redshifts for J0256, the GMM $\log(L)$ test is not a strong discriminator between the two models. However, based on the following additional evidence, we adopt the bimodal model in further analysis of this cluster.

Firstly, there are two BCGs (cluster members with the lowest SDSS magnitudes) that are spatially separated, as seen from the SDSS image in Figure \ref{fig:opticalmembers}, which are also separated in velocity space as shown in Figure \ref{fig:zvhisto}. This provides support for the existence of two distinct galaxy populations. These galaxies coincide with the peaks in the \emph{XMM-Newton} X-ray emission (see Figure \ref{fig:GMRT610}). Secondly, the DS test, which measures the deviation of the velocity distribution in spatially localised regions of a cluster with respect to the cluster as a whole, indicates the presence of substructure in J0256, with $S_{\Delta} < 0.01$ at the 68\% confidence level \citep{Sifon.2015}.

We use the GMM code to provide, for each member galaxy, the probability that the galaxy belongs to each of the kinematic components in the multi-variance bimodal case. In the following section we use these probabilities to calculate physical properties for the cluster and its components.

\subsubsection{Velocity dispersions and dynamical masses}
\label{sec:optmass}
By fitting a 2-mode GMM to our data, each cluster member is assigned a probability of belonging to each of the modes. These probabilities can be used to determine the mean and variance for each mode by integrating over all members and weighting by the probabilities. Since we have a discrete number of member galaxies, the mean and variance for component $n$ are given by
\begin{equation}
 \bar{z}_n = \left<z\right>_n = \frac{\sum_i p_n(z_i) z_i}{\sum_i p_n(z_i)} \label{eqn:popmean}
\end{equation}
\begin{equation}
 \sigma^2_{z,n} = \left<(z - \bar{z})^2\right>_n = \frac{\sum_i p_n(z_i) z_i^2}{\sum_i p_n(z_i)} - \left<z\right>_n^2 \label{eqn:popvar}
\end{equation}
\noindent where $n \in \left\lbrace{1, 2}\right\rbrace$, $z_i$ is the redshift of the $i$-th member galaxy, and $p_n(z_i)$ is the probability that this member belongs to the $n$-th component. The mean and variance of each mode in the redshift distribution correspond to the peak redshift and velocity dispersion for each kinematic component, respectively. We use the velocity dispersion and the galaxies-based scaling relation from \citet{Munari.2013.massprofile} to determine $M_{200}$ and $R_{200}$ for each component\footnote{$M_{200} = (4\pi/3) \rho_{200} R_{200}^3$}, using a value of $h = 0.7$ in the \citet{Munari.2013.massprofile} equation. Using the concentration parameter from \cite{Duffy.2008.DMHalosWMAP5}, we integrate a NFW profile \citep{Navarro.1997.NFWProfile} and interpolate to determine $M_{500}$ and $R_{500}$. The results are given in Table \ref{table:optclustcomps}, with all uncertainties determined via bootstrapping. We follow the same process using the unimodal fit, the difference being that 
the probability for every member is 1. 

\begin{table*}
 \centering
 \caption{Optical statistics of the two cluster components from 78 spectroscopic galaxy redshifts. $v_{\rm pec}$ is relative to $z$ = 0.363.}
 \begin{tabular*}{\textwidth}[h]{l @{\extracolsep{\fill}} cccccccc}
    \hline\hline
    Component & No. of galaxies & $z_{\rm mean}$  & $v_{\rm pec}$ & $\sigma$ & $M_{200}$ & $M_{500}$ & $R_{200}$ & $R_{500}$\\
     & & & (km s\per) & (km s\per) & ($10^{14} M_\odot$) & ($10^{14} M_\odot$) & (Mpc) & (Mpc)\\
    \hline
    main cluster & 59 & 0.361 $\pm$ 0.001 & -490 $\pm$ 100 & 850 $\pm$ 70 & 4.90 $\pm$ 1.03 & 3.23 $\pm$ 0.66 & 1.45 $\pm$ 0.11 & 0.92 $\pm$ 0.06 \\
    
    subcluster & 19 & 0.369 $\pm$ 0.002 & 1390 $\pm$ 180 & 690 $\pm$ 120 & 2.76 $\pm$ 1.14 & 1.83 $\pm$ 0.74 & 1.20 $\pm$ 0.19 & 0.76 $\pm$ 0.12\\
    \hline
  \end{tabular*}
  \label{table:optclustcomps}
\end{table*}

From the mean redshifts of the components, we find a line-of-sight velocity difference of $v_\perp = 1880 \pm 210$ km s\per. We also calculate individual component masses of $M_{\rm 500,main} = (3.23 \pm 0.66) \times 10^{14} M_\odot$ and $M_{\rm 500,subcl.} = (1.83 \pm 0.74) \times 10^{14} M_\odot$, leading to a merger mass ratio of 7:4, smaller than but within the errors of the $\sim$ 3:1 ratio determined by M04. Combining the component masses, we calculate a cluster dynamical mass of $M_{\rm 500,opt} = (5.06 \pm 0.99) \times 10^{14} M_\odot$, which agrees with the SZ cluster mass given in Table \ref{table:j0256} to better than 0.5$\sigma$. The combined $M_{\rm 200, opt}$ mass, $M_{\rm 200,opt} = (7.66 \pm 1.54) \times 10^{14} M_\odot$, agrees within 1$\sigma$ with the estimated $M_{\rm 200,X}$ total cluster mass range from M04 of $M_{\rm 200,X} \approx 9.7 - 11.1 \times 10^{14} M_\odot$, assuming a 15\% uncertainty on their $M_{200,NE}$ value\footnote{Corrected for the cosmology in this paper.}. 

If we model the cluster as a single component, we estimate a total mass $M_{\rm 500,tot} = (7.74 \pm 0.02) \times 10^{14} M_\odot$, which is 2.3$\sigma$ away from the SZ mass. This reinforces our argument in favour of the bimodal model. The corresponding $M_{200}$ measurement, $M_{\rm 200,tot} \sim 11.7 \times 10^{14} M_\odot$, still agrees with the total X-ray mass estimate from M04, although this comparison is not particularly meaningful given the large uncertainties on their estimate.

\section{Merger geometry}
\label{sec:mergergeom}
M04 construct a simple merger model for J0256 using projected distances and the line-of-sight velocity difference between the main and subcluster components. We adopt a similar approach but update two aspects: we use a more current cosmology and the increased number of galaxy spectroscopic redshifts (78 vs. 4) discussed in Section \ref{sec:arch-Opt}. The optical galaxy redshift distribution also allows us to determine dynamical masses for the main and subcluster components.

\begin{figure}
 \centering
 \includegraphics[width=0.45\textwidth]{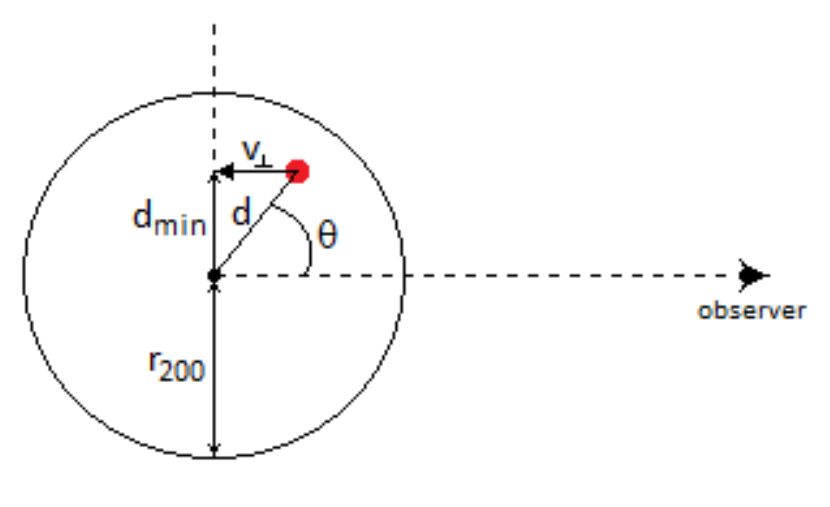}
 \caption{Merger geometry of J0256 as per \citet{Majerowicz.2004.J0256}. The small black dot represents the centre of the main cluster component and the red circle represents the centre of the subcluster. $d_{\rm min}$ and $d$ are the projected distance and physical distance between the two component centres, respectively. $v_\perp$ is the line-of-sight infall velocity and $\theta$ is the impact angle.}
 \label{fig:geometry}
\end{figure}

\begin{figure}
 \centering
 \includegraphics[width=0.5\textwidth,clip,trim=0 0 0 0]{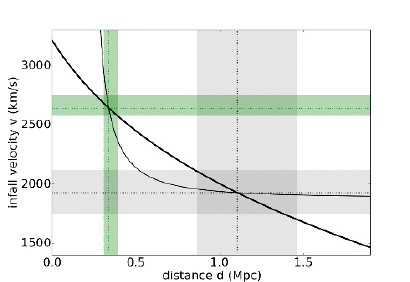}
 \caption{Trigonometric (solid, thick; eqn. \ref{eqn:vdrel2}) and integrated NFW profile (solid, thin; eqn. \ref{eqn:vdrel1}) relations between infall velocity $v$ and cluster component separation $d$. The intersections of the two relations give the two possible solutions for $v$ and $d$. The grey dotted lines and shaded regions indicate the solution for case one: $v_1 = 1930^{+190}_{-170}$ km s{\per} and $d_1 = 1.105^{+0.353}_{0.241}$ Mpc. The green dotted lines and shaded areas indicate the solution for case two: $v_2 = 2640^{+110}_{-60}$ km s{\per} and $d_2 = 0.338^{+0.056}_{-0.020}$ Mpc. }
 \label{fig:vdrelation}
\end{figure}

\begin{table*}
 \centering
 \caption{Merger geometry and time-scales from today for two possible cases with $d_{\rm min}$ = 237.6 kpc and $v_\perp$ = 1880 {$\pm$} 210 km s\per.}
  \begin{tabular}[h]{lcccccccccccccc}
    \hline\hline
    
     & & $v$ & & $d$ & & $\theta$ & & $-t_{\rm A}$ $^a$ & & $t_{\rm B}$ $^b$ & & $t_{\rm C}$ $^c$ & & $\Gamma$ $^d$\\
     & & (km s\per) & & (kpc) & & (degrees) & & (Gyr) & & (Gyr) & & (Gyr) & & (\%) \\
    \hline
    case 1 & & $1930^{+190}_{-170}$ & & $1110^{+350}_{-240}$ & & $12^{+5}_{-3}$ & & $1.06^{+0.23}_{-0.20}$ & & $0.46^{+0.21}_{-0.22}$ & & $1.99^{+0.20}_{-0.22}$ & & $35 ^{+7}_{-18}$ \vspace*{0.2cm}\\

    case 2 & & $2640^{+110}_{-60}$ & & $340^{+60}_{-20}$ & & $45^{+6}_{-7}$ & & $1.41^{+0.03}_{-0.04}$ & & $0.12^{+0.03}_{-0.03}$ & & $1.64^{+0.03}_{-0.03}$ & & $46 ^{+1}_{-2}$ \\
    \hline
  \end{tabular}
  \justify
  $^a$ Time since first virial crossing.\\
  $^b$ Time until core passage.\\
  $^c$ Time until second virial crossing.\\
  $^d$ Measure of how far along in the merger the cluster currently is, $\Gamma = |t_{\rm A}/t_{\rm tot}| = |t_{\rm A}/(t_{\rm C} - t_{\rm A})|$.  
  \label{table:mergerstats}
\end{table*}

\begin{figure}
 \centering
 \includegraphics[width=0.5\textwidth,clip,trim=0 0 -5 0]{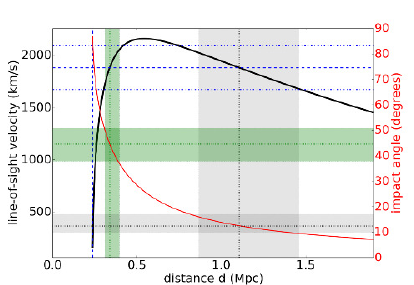}
 \caption{The line-of-sight velocity $v_\perp$ versus the component separation $d$ (black, thick, solid) using the relation in equation \ref{eqn:vdrel2} with the infall velocity $v(d)$ given by equation \ref{eqn:vdrel1}. The solid thin red curve is the impact angle $\theta$ as a function of $d$. The vertical and horizontal blue dashed lines indicate the values of $d_{\rm min} = 237.6$ kpc and $v_\perp = 1880 \pm 210$ km s{\per} respectively, with the horizontal blue dot-dashed lines indicating the lower and upper limits for $v_\perp$. Figure \ref{fig:vdrelation} showed the two merger geometry solutions. Here the grey dotted lines and shaded areas indicate the $d$ and $\theta$ values and uncertainties for case one: $d_1 = 1.11^{+0.35}_{0.24}$ Mpc and $\theta_1 = 12$\degrees$^{+5}_{-3}$. The green dotted lines and shaded regions indicate the same, but for case two: $d_2 = 0.34^{+0.06}_{-0.02}$ Mpc and $\theta_2 = 45$\degrees$^{+6}_{-7}$.}
 \label{fig:vperp-d-theta}
\end{figure}

For simplicity, we assume the same merger geometry as in M04, schematically outlined in Figure \ref{fig:geometry}. Working in the rest frame of the main component, we assume the same simplification of a point mass subcluster and ignore dynamical friction. However, rather than using a $\beta$-model, we assume the mass distribution of the main component is defined by a NFW profile \citep{Navarro.1997.NFWProfile}: 
\begin{equation}
 M(<R) = {4\pi\rho_0} R_s^3 \left[\ln(1+c) - \frac{c}{1 + c}\right]
 \label{eqn:NFW}
\end{equation}
\noindent where $R_s = R/c$ is a characteristic scale radius, $c$ is the concentration parameter for radius $R$, and $\rho_0$ is the typical NFW dark matter density for the cluster. Using the $c(M,z)$ relation from \citet{Duffy.2008.DMHalosWMAP5} to determine $c$ for our cluster, we have $c = 3.018$ and $\rho_0 = 5.497 \times 10^{14} M_\odot$Mpc$^{-3}$.

Using the above mass profile and modelling the gravitational infall of the subcluster, we obtain the following relation between subcluster infall velocity, $v$, and physical separation, $d$, between the centres of the subcluster and the main component:

\begin{equation}
 v^2(d) = \frac{2 G M_{200}}{R_{200}} + \frac{2 G M_{0}}{R_s} \left[\frac{\ln(1+d/R_s)}{d/R_s} - \frac{\ln(1+c)}{c}\right],
 \label{eqn:vdrel1}
\end{equation}

\noindent where $M_0 = 4 \pi \rho_0 R_s^3$. The subcluster redshift $z_{\rm sub}$ is greater than that of the main cluster component, $z_{\rm main}$. As we argued in Section \ref{sec:arch-Xray}, the X-ray emission pattern indicates that the subcluster is moving towards the main component. This implies that the impact angle must be less than 90\degrees. Using simple trigonometry, it follows from the merger geometry in Figure \ref{fig:geometry} that 
\begin{equation}
 \left(\frac{d_{\rm min}}{d}\right)^2 + \left(\frac{v_\perp}{v}\right)^2 = 1.
 \label{eqn:vdrel2}
\end{equation}
\noindent where $d_{\rm min}$ is the projected separation between the main component and the subcluster, and $v_\perp$ is the velocity difference along the line-of-sight.

Using the X-ray peaks of each component, the projected separation between cluster components is $\sim$0.78{\arcmin}, which corresponds to a physical projected distance of $d_{\rm min} = 237.6$ kpc (as compared to 350 kpc in M04). In Section \ref{sec:GMM} we found $v_\perp = 1880 \pm 210$ km s{\per} which is consistent with the value estimated by M04. Based on the X-ray arguments in Section \ref{sec:arch-Xray}, the two cluster components have begun interacting and we can place the following limits on the physical separation and the infall velocity: $d_{\rm min} < d < R_{200}$ and $v > v_\perp$, where $R_{200}$ is the cluster radius for the main component. 

Simultaneously solving equations \ref{eqn:vdrel1} and \ref{eqn:vdrel2} with these constraints provides two sets of solutions for the merger model. These are listed in Table \ref{table:mergerstats}, with the graphical solutions given in Figures \ref{fig:vdrelation} and \ref{fig:vperp-d-theta}. The uncertainties on $v$, $d$, and $\theta$ are shown in Figures \ref{fig:vdrelation} ($v$ and $d$) and \ref{fig:vperp-d-theta} ($d$ and $\theta$), and are propagated from the uncertainties on the $R_{200}$ mass and radius of the main cluster component, the $R_{200}$ radius of the subcluster, and measured line-of-sight velocity difference. We consider these solutions in the next section to estimate relevant time-scales in the merger.

\section{Merger and radio halo time-scales}
\label{sec:timescales}
To better understand the formation history and mechanism(s) of GRHs, we would like to relate the GRH formation time-scale to the merger time-scale. It is possible to model the physics of turbulent re-acceleration using simulations. \citet{Donnert.2013.MHDSims} (hereafter D13) used MHD simulations of a $10^{15} M_\odot$ and 8:1 merger to study the strength and pattern of diffuse radio emission at various merger stages. They found that the cluster needs to have been actively merging for a minimum amount of time, approximately 15\% into the merger, such that there is sufficient turbulence generated, before the radio emission switches on.

\subsection{Estimates for merger time-scales}
\label{sec:mergertimes}
\begin{figure*}
 \centering
 \includegraphics[width=0.8\textwidth,clip,trim=100 20 100 50]{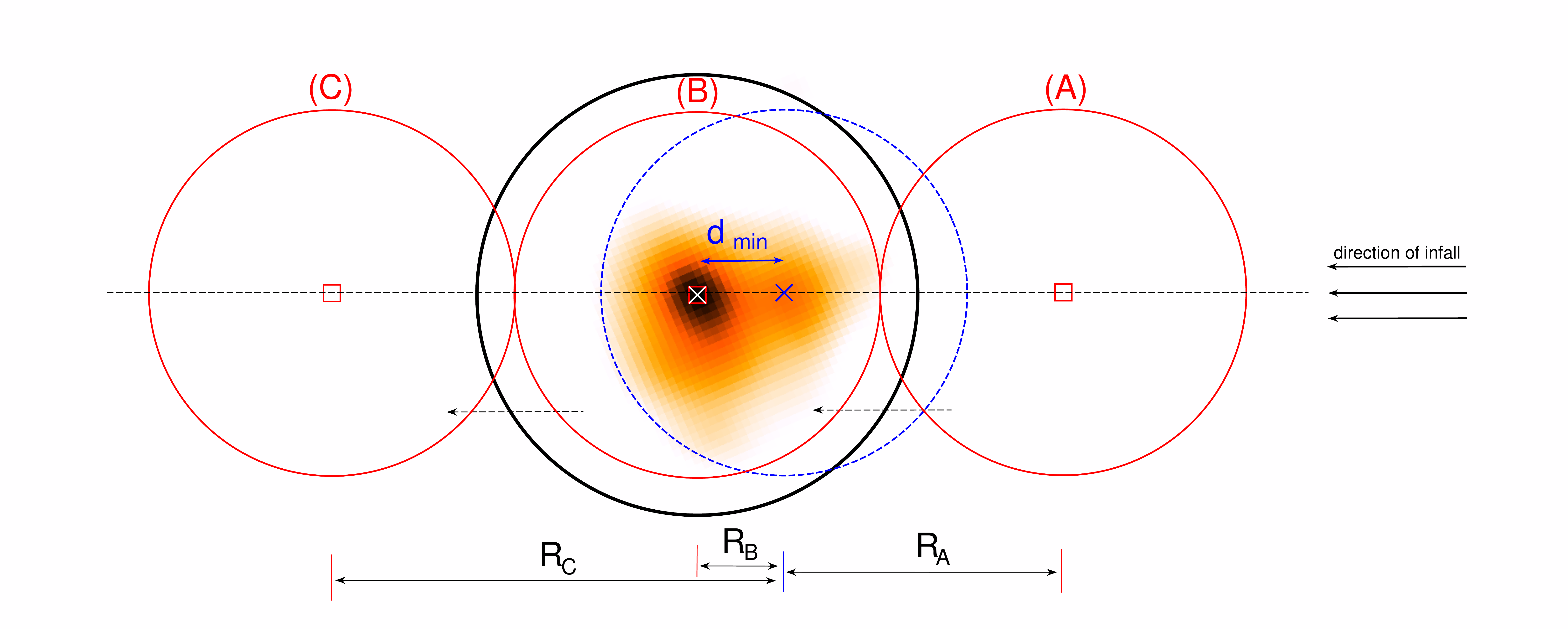}
 \caption{Schematic showing the relative position of the subcluster (red circles) to the main cluster (black circle) overlaid on the X-ray image at three different times during the merger: (A) first virial crossing; (B) core passage; and (C) second virial crossing. The centre of the main cluster is marked with a white cross while the centre of the subcluster at each interval of the merger is shown by a red diamond. The blue cross and dashed circle denotes the current position of the subcluster. The dashed black line represents the merger axis and $d_{\rm min}$ is the projected distance between the two cluster components. All circles denote $R_{200}$ of the respective components. Even though we refer to virial crossing, we use $R_{200}$ as a proxy for the virial radius.}
 \label{fig:mergertimes}
\end{figure*}

To estimate the merger time-scales for J0256 we assume a simple merger taking place in a linear fashion along the merger axis determined by the impact angle, $\theta$, schematically outlined in Figure \ref{fig:mergertimes}. In Section \ref{sec:arch-Xray}, we ruled out a scenario in which the subcluster has already passed through the core. In Figure \ref{fig:mergertimes}, we isolate three distinct times during the merger: (A) first virial crossing; (B) core passage; and (C) second virial crossing. Even though we refer to virial crossing, we use $R_{200}$ as a proxy for the virial radius.

From the optical analysis in Section \ref{sec:optical}, $R_{200}^{\rm main}$ = 1.45 Mpc and $R_{200}^{\rm subcl}$ = 1.20 Mpc. First virial crossing thus occurs when the centres of the two components are initially 2.65 Mpc apart. The distances associated with the three merger stages are $R_A = 2.65 \text{ Mpc} - d$, $R_B = d$, and $R_C = 2.65 \text{ Mpc} + d$, where $d$ is the current physical separation for the two model solutions listed in Table \ref{table:mergerstats}. 

For each merger model solution found in the previous section, we compute the merger times 
\begin{equation}
 t_n = \int^{R_n}_{R_0}\frac{\text{d}R^\prime}{v_{_{\rm NFW}}(R^\prime)}
\end{equation}

\noindent where $n \in \left\lbrace {\rm A, B, C} \right\rbrace$, $R_0$ is the observed position of the subcluster, and $v_{_{\rm NFW}}$ is the velocity function given in equation \ref{eqn:vdrel1}. The total time of the merger, at least for the first passage, is given by $t_{\rm tot} = t_{\rm C} - t_{\rm A}$. We define the relative time phase of the merger as the ratio $\Gamma = |t_{\rm A}/t_{\rm tot}|$. The results for each model solution are given in Table \ref{table:mergerstats}. 

For case 2, we find that J0256 would have completed first virial crossing 1.41 Gyr ago with $\sim$120 Myr until first core passage occurs. This puts the cluster $\Gamma_2 = 46^{+1}_{-2}$\% of the way into its merger. In case 1, J0256 is closer to the beginning of its merger with $\sim$460 Myr until first core passage. The time-scales for case 1 result in J0256 having a relative time phase of $\Gamma_1 = 35^{+7}_{-18}$\%. According to D13, these conclusions lead to very different theoretical predictions for the observed strength and morphology of the radio emission. In the following section we compare our time-scale results with the D13 simulations.

\subsection{Comparison with MHD simulations}
\label{sec:radiopredictions}
The simulated radio powers and morphologies in D13 are for observations at 1.4 GHz of a massive $10^{15} M_\odot$ cluster undergoing a 8:1 mass ratio, plane-of-the-sky merger. J0256 is about 50\% of the total simulated mass but has a much smaller mass ratio of 7:4. As the strength, and hence observability, of the radio emission is related to cluster mass and the amount of turbulent energy created during a merger, we caution that, for the specific case of J0256, the following comparison with the D13 results can at best be qualitative due to the above differences between J0256 and the simulated cluster. MHD simulations for the particular case of J0256 would be required for a more accurate comparison.

To compare our merger time-scales with the MHD simulations of D13 we need to convert our values into their time frame. From the X-ray snapshots of their simulated merger (see their Figure 3), we estimate first and second virial crossings to occur at 0 Gyr and 2.56 Gyr respectively, giving $t_{\rm tot,D13} = 2.56$ Gyr, similar to the total merger time of 3.05 Gyr for J0256. Scaling our $\Gamma$ values to this time-scale allows us to extrapolate expected radio power and general emission morphology for each case in Table \ref{table:mergerstats} using the D13 simulation.

In case 1 we have $\Gamma_1 = 35^{+7}_{-18}$\%, corresponding to $t_{\rm A,D13} = 0.90^{+0.18}_{-0.46}$ Gyr. Here, not enough turbulence is being generated to drive the diffuse radio emission and only compact radio source emission is observable in Figure 3 of D13. Case 2 gives $\Gamma_2 = 46^{+1}_{-2}$\%, corresponding to $t_{\rm A,D13} = 1.18^{+0.02}_{-0.05}$ Gyr. Here the GRH is in the early stages of having switched on, according to Figure 3 of D13, and is gaining power. The X-ray image of J0256 shown in Figure \ref{fig:xmmepic} is a close visual match with the second panel of Figure 3 in D13, which has a relative time-scale similar to that of case 2. This consistency is in contrast to case 1, where no diffuse radio emission is observable and the expected radio power lies in the realm of the upper limits on the $P_{\rm 1.4GHz}$--$L_{\rm X}$ scaling relation. Thus our case 2 appears to be the more likely of the two merger geometry solutions for J0256: we observe what is likely a young radio halo. 

\section{Conclusion}
\label{sec:conclusion}
We have detected a low surface brightness giant radio halo ($\sim$0.8 Mpc) in ACT-CL J0256.5+0006 with the GMRT at 610 MHz, and obtained a marginal detection at 325 MHz. With an SZ mass of $M_{500} = (5.0 \pm 1.2) \times 10^{14} M_\odot$, J0256 is one of the lowest mass clusters currently known to host such emission. 

We measure halo flux densities of $S_{610} = 5.6 \pm 1.4$ mJy and $S_{325} = 10.3 \pm 5.3$ mJy, giving a measured spectral index of $\alpha^{610}_{325} = 1.0^{+0.7}_{-0.9}$. Due to the unreliability of the 325 MHz measurements, we calculate a bandwidth- and k-corrected 1.4 GHz radio power of $P_{1.4\text{GHz}} = (1.0 \pm 0.3) \times 10^{24}$ W Hz{\per} by extrapolating our 610 MHz flux density to 1.4 GHz using a theoretically motivated spectral index of $\alpha = 1.2 \pm 0.2$. As the detection at 610 MHz is not highly significant, we do not draw strong conclusions about the radio morphology, but we do note that it roughly follows the thermal gas as seen in the X-rays and is centred on the cluster SZ peak. More data at 325 MHz would be required to confirm our detection at this frequency and obtain a more accurate measured spectral index.

Using the X-ray and optical information available to us, we have investigated the morphology of J0256, concluding that this system consists of a main cluster component with an in-falling subcluster slightly in front and to the west of it. The merger mass ratio determined via new spectroscopic galaxy member redshifts is roughly 7:4, making it a major merger event. We estimate a line-of-sight velocity difference between the two components of $v_\perp = 1880 \pm 210$ km s{\per}. 

Using this information and assuming an NFW mass profile and a simple merger geometry defined by $v$, $d$, and $\theta$, we find two possible solutions for the merger time-scale. Defining the merger time phase, $\Gamma$, to be the percentage of the first passage (between first and second virial crossings) already completed, we find that J0256 has a merger time phase of $\Gamma_2 = 46^{+1}_{-2}$\% or $\Gamma_1 = 35^{+7}_{-18}$\%. We compare these values with MHD simulations from \citet{Donnert.2013.MHDSims} and conclude that J0256 is most likely $\sim$47\% of the way into its merger, with only $\sim$100 Myr until first core passage. As the strength of the synchrotron emission is related to the amount of turbulent energy produced during a merger, a population of simulations varying in cluster mass and merger ratio would be useful in investigating the GRH formation rate for a wider range of models.

Our discovery of a GRH in J0256 may help to provide some insight into whether GRHs exist in all merging clusters and whether the non-detections in known merging systems are due to a combination of a low-mass cluster and insufficient sensitivity to diffuse emission, rather than to a complete lack of GRHs. More systems like J0256 will probe the full evolving population of GRHs, in particular the early-stage mergers, and potentially fill in the gap between radio upper limits and USSRHs in the $P_{\rm 1.4GHz}$--$L_{\rm X}$ plane. It would be interesting to carry out a similar merger time-scale analysis for existing GRHs to probe the scatter in the radio power scaling relations.

\section*{Acknowledgements}
The authors thank the anonymous referee whose comments have greatly improved the manuscript, and G. Brunetti for useful comments on the original arXiv version.

KK acknowledges post-graduate support from the NRF/SKA South Africa Project. HTI is financially supported by the National Radio Astronomy Observatory, a facility of the National Science Foundation operated under Associated Universities Inc. AJB acknowledges support from National Science Foundation grant AST-0955810.

We thank the staff of the GMRT that made these observations possible, and the Director for approving DDT. GMRT is run by the National Centre for Radio Astrophysics of the Tata Institute of Fundamental Research. Results in this paper are based on observations obtained at the Gemini Observatory (ObsID:GS-2011B-C-1, GS-2012A-C-1), which is operated by the Association of Universities for Research in Astronomy, Inc., under a cooperative agreement with the NSF on behalf of the Gemini partnership: the National Science Foundation (United States), the National Research Council (Canada), CONICYT (Chile), the Australian Research Council (Australia), Minist\'{e}rio da Ci\^{e}ncia, Tecnologia e Inova\c{c}\~{a}o (Brazil) and Ministerio de Ciencia, Tecnolog\'{i}a e Innovaci\'{o}n Productiva (Argentina).

\appendix
\section{Full-resolution and Low-Resolution Radio Maps}
In this appendix we provide the inner 30{\arcmin} $\times$ 30{\arcmin} of the full resolution and smoothed low resolution maps for both 610 MHz and 325 MHz. In each image, the dashed circle indicates the cluster scale $\theta_{500} = 3.1${\arcmin} from \citet{Hasselfield.2013.ACTE}, centred on the SZ cluster peak, which is shown as a red or white X. The solid circle shows the {13\arcmin} radius outside of which we removed all compact emission before further imaging in CASA, as described in Section \ref{sec:radioobs}.

\bibliography{references}

\begin{figure*}
  \centering
  \includegraphics[width=0.58\textwidth, trim=200 20 200 20]{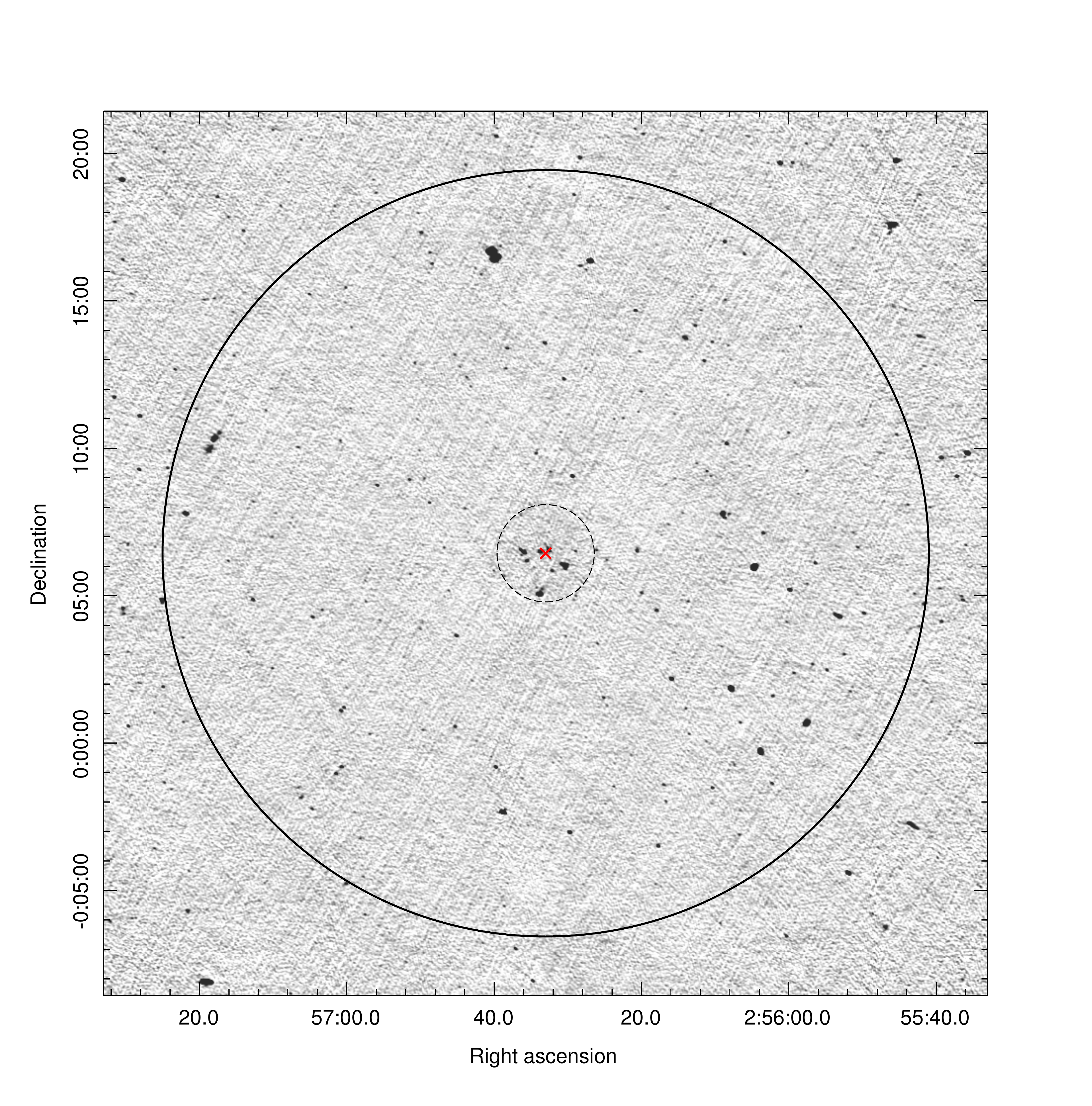}
  \caption{Inner 30{\arcmin} $\times$ 30{\arcmin} of the full-resolution (FR) 610 MHz map. The beam is 5.7{\arcsec} $\times$ 4.1{\arcsec} at p.a. 71.3{\degrees}, and the map noise is $\sigma$ = 26 $\mu$Jy beam\per. The dashed black circle represents $\theta_{500} = 3.1${\arcmin}, centred on the cluster SZ peak shown by the red X. The 13{\arcmin} radius is shown by the solid black circle.}
  \label{fig:610BEST}
 
\end{figure*}

\begin{figure*}
 \centering
 \includegraphics[width=\textwidth,trim=400 20 400 20]{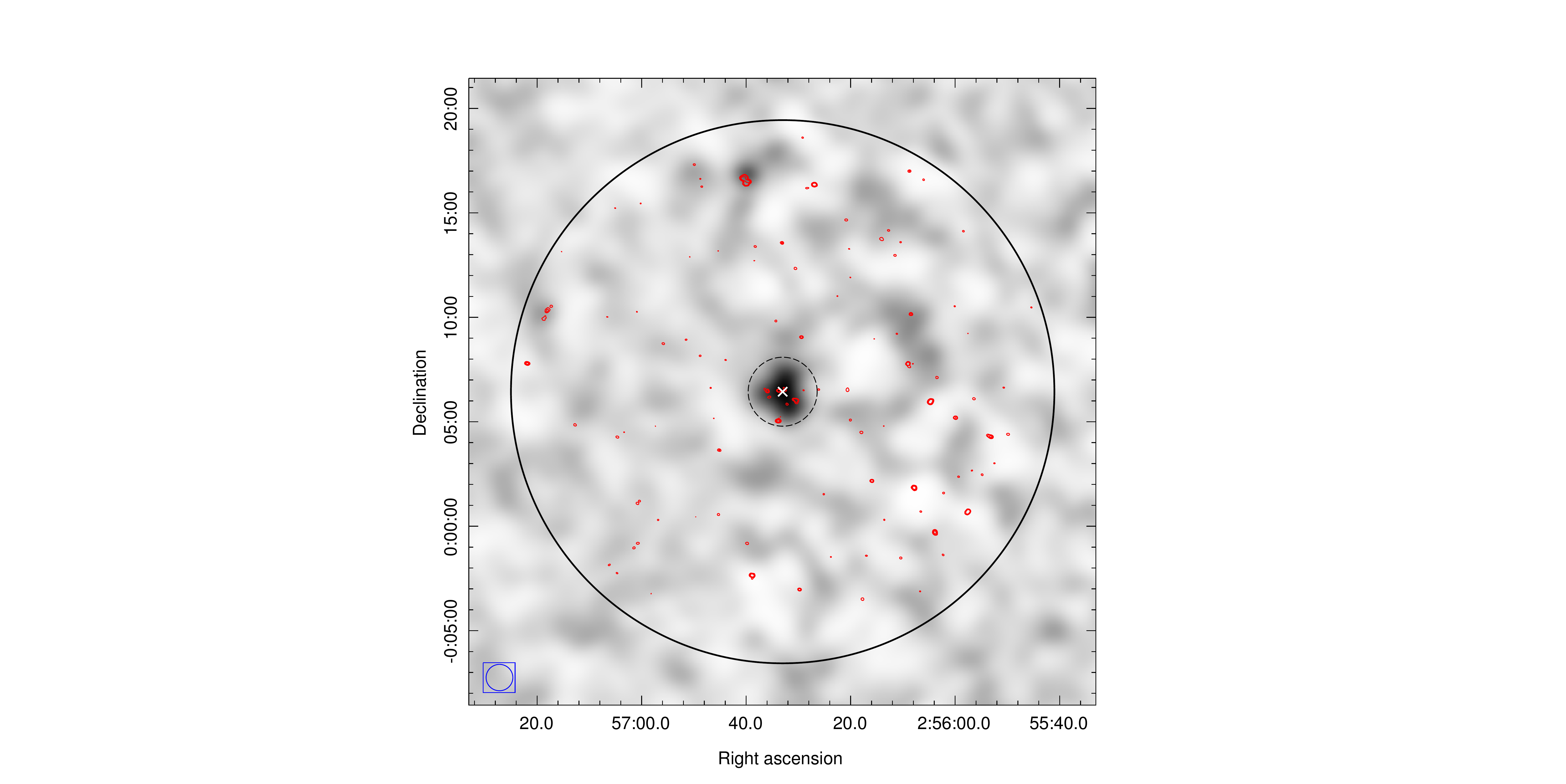}
 \caption{Inner 30{\arcmin} $\times$ 30{\arcmin} of the 610 MHz map. Greyscale is the low-resolution (LR), 1{\arcmin}-smoothed image. Red contours are the high-resolution (HR) [6, 20, 80]$\times 1\sigma$ contours where $1\sigma$ = 31 $\mu$Jy beam\per. The X and black solid and dashed circles are as in Figure \ref{fig:610BEST}. The LR beam is 79.6{\arcsec} $\times$ 76.8{\arcsec} at p.a. -86.9{\degrees} and is shown by the blue ellipse in the lower left corner. The 1$\sigma$ noise in the LR greyscale image is 0.36 mJy beam{\per}.}
 \label{fig:610LR}
\end{figure*}

\begin{figure*}
 \centering
 \includegraphics[width=0.58\textwidth, trim=200 20 200 20]{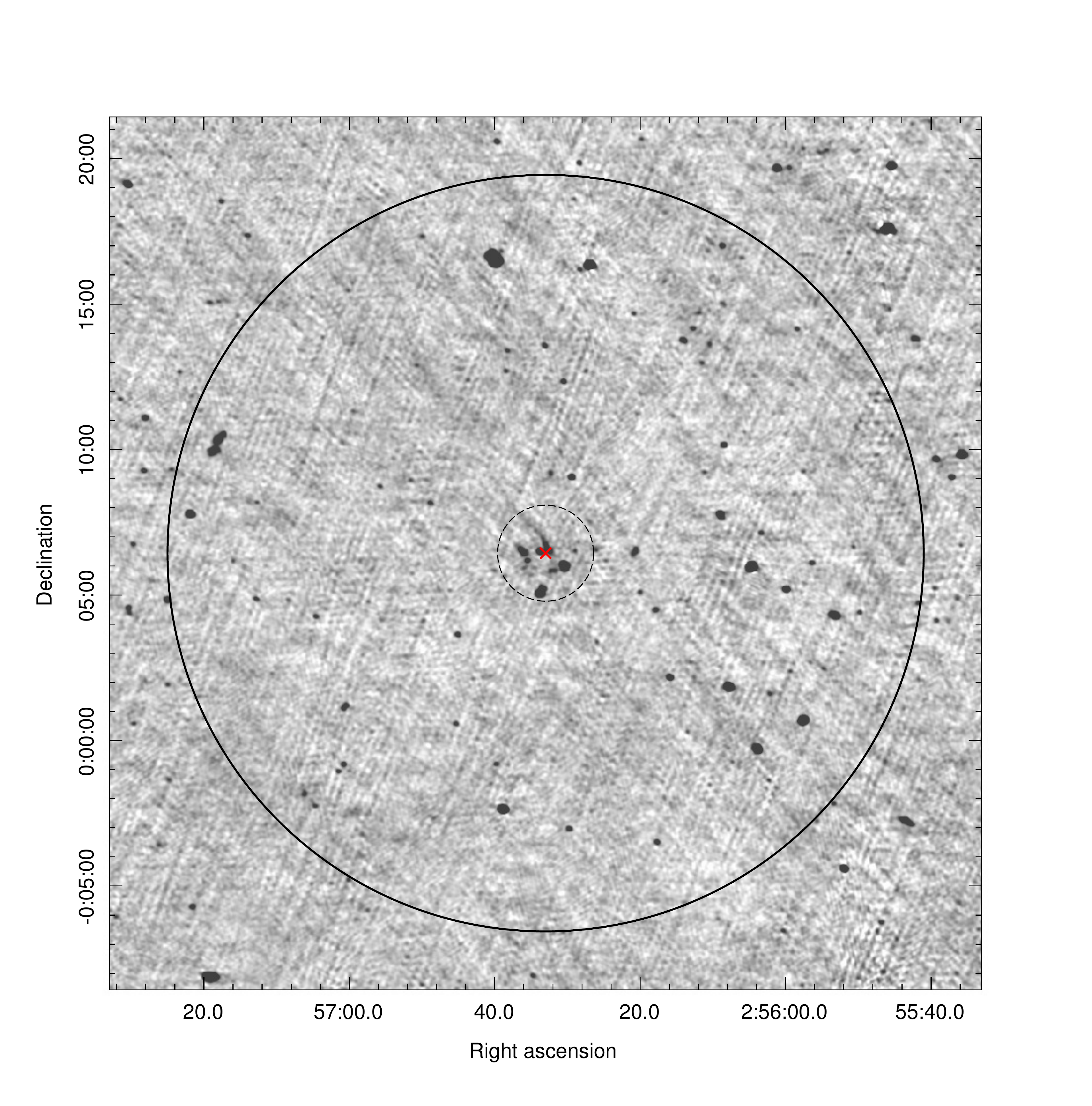}
 \caption{Inner 30{\arcmin} $\times$ 30{\arcmin} of the full-resolution (FR) 325 MHz map. The beam is 9.7{\arcsec} $\times$ 7.9{\arcsec} at p.a. 74.1{\degrees} and the map noise is $\sigma$ = 77 $\mu$Jy beam\per. The X and black solid and dashed circles are as in Figure \ref{fig:610BEST}.}
 \label{fig:330BEST}
\end{figure*}

\begin{figure*}
 \centering
 \includegraphics[width=0.58\textwidth, trim=200 20 200 20]{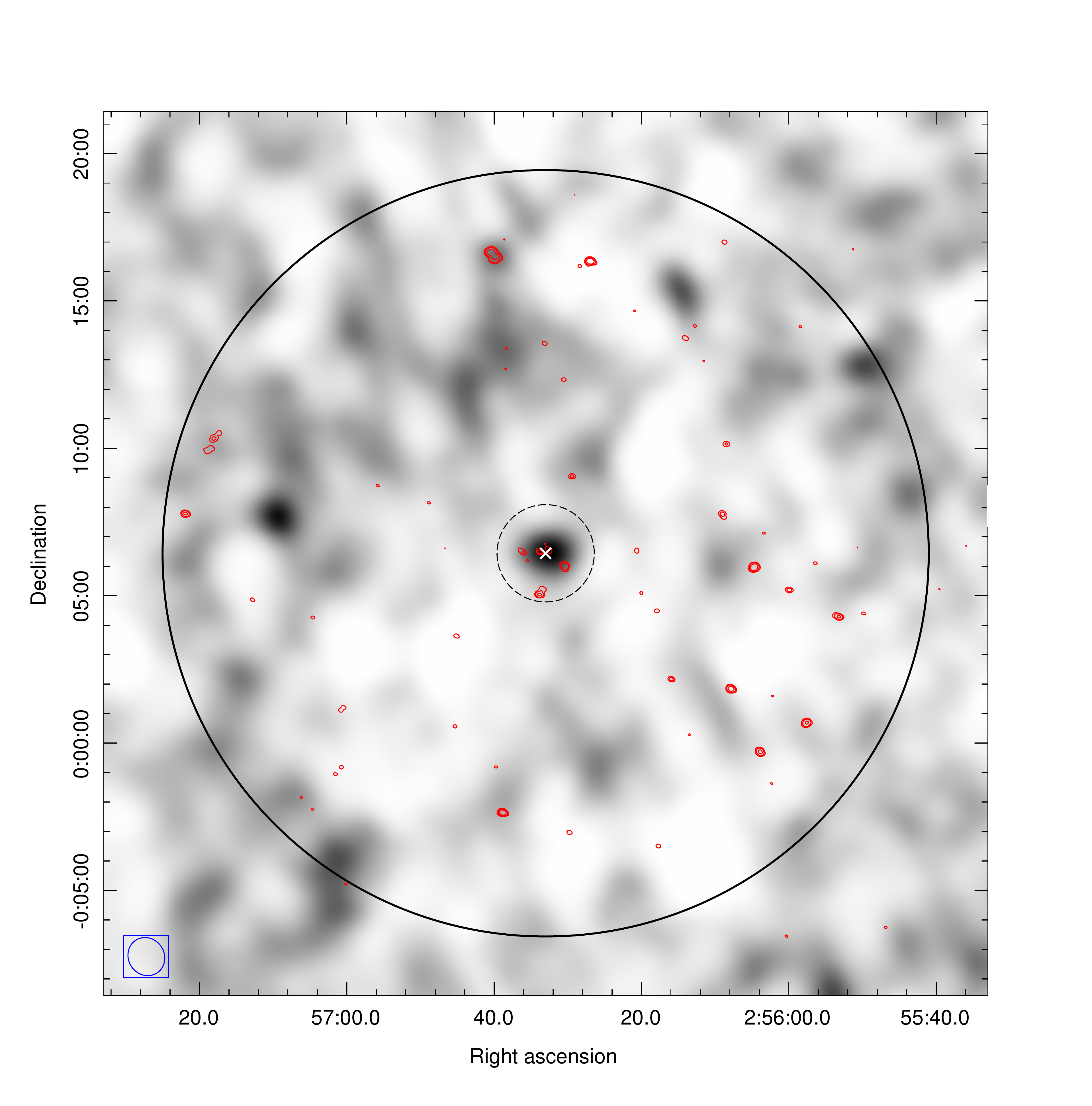}
 \caption{Inner 30{\arcmin} $\times$ 30{\arcmin} of the 325 MHz map. Greyscale is the low-resolution (LR), 1{\arcmin}-smoothed image. Red contours are the high-resolution (HR) [6, 20, 80]$\times 1\sigma$ contours where $1\sigma$ = 71 $\mu$Jy beam\per. The X and black solid and dashed circles are as in Figure \ref{fig:610BEST}. The LR beam is 79.4{\arcsec} $\times$ 73.1{\arcsec} at p.a. 56.7{\degrees} and is shown by the blue ellipse in the lower left corner. The 1$\sigma$ noise in the LR greyscale image is 1.18 mJy beam{\per}.}
 \label{fig:330LR}
\end{figure*}

% Don't change these lines
\bsp	% typesetting comment
\label{lastpage}
\end{document}